\documentclass[pra,twocolumn,prl,groupedaddress]{revtex4-1}

\usepackage{graphicx}
\usepackage{dcolumn}
\usepackage{bm}
\usepackage{epsfig}
\usepackage{amsmath}
\usepackage{amssymb}
\usepackage{color}
\usepackage{bbm}
\usepackage{braket}

\usepackage[colorlinks=true,citecolor=blue,linkcolor=blue,urlcolor=blue]{hyperref}

\begin{document}

\title{Cavity quantum electrodynamics in the non-perturbative regime}

\author{Daniele De Bernardis, Tuomas Jaako, and Peter~Rabl}
\affiliation{Vienna Center for Quantum Science and Technology, Atominstitut, TU Wien, 1040 Vienna, Austria}

\date{\today}

\begin{abstract} 
We study a generic cavity-QED system where a set of (artificial) two-level dipoles is coupled to the electric field of a single-mode $LC$ resonator. This setup is used to derive a minimal quantum mechanical model for cavity QED, which accounts for both dipole-field and direct dipole-dipole interactions. The model is applicable for arbitrary coupling strengths and allows us to extend the usual Dicke model into the non-perturbative regime of QED, where the dipole-field interaction can be associated with an effective finestructure constant of order unity. In this regime, we identify  three distinct classes of normal, superradiant and subradiant vacuum states and discuss their characteristic properties and the transitions between them. Our findings reconcile many of the previous, often contradictory predictions in this field and establish a common theoretical framework to describe ultrastrong coupling phenomena in a diverse range of cavity-QED platforms. 
\end{abstract}
 
\maketitle

%
%


\maketitle

\section{Introduction}
Quantum electrodynamics (QED) is the fundamental theory of charges and electromagnetic fields, which in its low-energy limit describes the physics of photons interacting with atoms, molecules and solids. Cavity QED~\cite{HarocheCQED} is a minimal framework within which such light-matter interactions are studied at the quantum level in terms of two-level emitters coupled to a single radiation mode. A hallmark of cavity QED is the strong coupling between single atoms and single photons, which  has been the subject of many theoretical and experimental works in this field. 
These strong interactions between excited atomic and photonic states are, however, still perturbative in the sense that the coupling is much smaller than the absolute atomic or photonic energy scales involved. Indeed, it follows from basic geometric considerations that the coupling strength $g$ between an elementary electric dipole and a cavity mode of frequency $\omega_c$  is limited to~\cite{HarocheCQED,Devoret2007,Comment}
\begin{equation}\label{eq:Bound}
\frac{g}{\omega_c}\lesssim  \sqrt{2\pi \alpha_{\rm fs}},
\end{equation}
where $\alpha_{\rm fs}\simeq1/137$ is the finestructure constant. As a consequence, the vacuum of (cavity) QED is to a good approximation represented by the trivial state with all atoms in their ground state and no photons. This is in stark contrast to the theory of quantum chromodynamics, where much more complex ground states of strongly interacting quarks and gluons arise. 

The interest in the physics of light-matter interactions beyond this `weak-coupling' regime dates back to the early days of cavity QED and is traditionally closely connected to the Dicke model~\cite{Dicke1954,Brandes2005,Garraway2011} describing the coupling of $N$ two-level atoms to a single optical mode. For a sufficiently strong \emph{collective} coupling, $G=g\sqrt{N}$, the ground state of this model undergoes a quantum phase transition from the normal vacuum into a so-called superradiant phase, where the atoms spontaneously polarize and the field acquires a non-vanishing expectation value~\cite{Hepp1973,Wang1973,Hioe1973}. Over the past decades the existence of such a cavity-induced instability has been subject of many controversial debates. Most notably, it has been argued~\cite{Rzazewski1975} that the superradiant phase does not occur in more realistic models when the usually neglected diamagnetic ``$A^2$-term'' is correctly taken into account. This famous no-go-theorem has both been confirmed as well as rejected by many subsequent studies of various cavity QED~\cite{Kudenko1975,Emelanov1976,Knight1978,Bialynicki1979,Bowden1979,Yamanoi1979,Keeling2007,Hagenmuller2010,Todorov2012,VukicsPRA2012,
DeLiberato2014,Vukics2014,Bamba2014,Vukics2015,Tufarelli2015,Griesser2016} and analogous circuit QED~\cite{CiutiNatComm2010,NatafPRL2010,ViehmannPRL2011,Leib2012,Jaako2016,Bamba2016,Bamba2017} setups, but despite its fundamental relevance, this matter is still not fully resolved.

More recently,  the development of various solid-state cavity QED platforms has led to a growing number of experimental activities related to what is now quite generally called the ultrastrong coupling (USC) regime~\cite{Ciuti2005} of light-matter interactions. By using, for example, organic materials~\cite{Schwartz2011,KenaCohen2013,Mazzeo2014}, intersubband transitions~\cite{Todorov2010,Geiser2012, Benz2013,Dietze2013,Vasanelli2016}, or 2D electron gases~\cite{Scalari2012,Maissen2014,Zhang2016,Keller2017}, the collective coupling of such dense dipolar ensembles to optical or THz modes can reach a considerable fraction of  the bare photon frequency. 
In parallel, it has been demonstrated in the context of circuit QED~\cite{Wallraff2004,Blais2004} that artificial atoms, like superconducting qubits~\cite{YouNature2011} or quantum dots \cite{Mi2017,Stockklauser2017,Bruhat2016,Cottet2017}, can be coupled very efficiently to microwave resonators, in which case the USC regime becomes accessible even at the single-qubit level~\cite{Forndiaz2017,Yoshihara2017,Niemczyk2010,Forndiaz2010,Baust2016, Chen2017,Bosman2017}. 
In light of these experimental developments and potential applications ranging from USC-assisted chemical reactions~\cite{Hutchison2012,Wang2014,Galego2015,Herrera2016,Flick2016} to ultra-fast superconducting quantum information processing schemes~\cite{Romero2012,Armata2017}, a refined understanding of the basic principles of USC cavity QED on the single-, few- and many-particle level becomes of uttermost importance.

In this work we analyze a generic cavity QED setup where multiple two-level dipoles are coupled to a single electromagnetic mode of a lumped-element $LC$ resonator. In this setup the limit on the interaction strength stated above can  be overcome by coupling (artificial) dipoles to the electric field of a tailored circuit mode with an impedance much higher than that of free space~\cite{Devoret2007}. In view of Eq.~\eqref{eq:Bound}, one can then associate with this system an effective finestructure constant of order unity, meaning that already for a single dipole a non-perturbative treatment of electromagnetic interactions must be taken into account.  The purpose of this study is, first of all, to derive a minimal consistent model for cavity QED, which is applicable in this non-perturbative regime~\cite{Comment2}, and second, to evaluate and describe the resulting vacuum states under various conditions.  
In contrast to most previous studies on this subject, we here focus explicitly on the long-wavelength and low-frequency regime to avoid many of the complications related to the quantization of the full electromagnetic field~
\cite{PhotonsAndAtoms,ScheelBook}. This approach still captures correctly the relevant low-energy physics and  allows us to rigorously separate the collective coupling to a single dynamical field mode from all direct dipole-dipole interactions.  Thereby, most of the ambiguities about the existence or non-existence of superradiant instabilities can be resolved and explained  in terms of basic electrostatic considerations. Our analysis also addresses several other subtle issues, like 
the breakdown of gauge invariance, which results in a unique extension of the Dicke model into the USC regime.

From the analysis of the ground states of this model we identify three distinct classes of  normal, superradiant and subradiant vacuum states,  which arise from the competition between direct dipole-dipole and cavity-mediated interactions. Our study first of all shows that a superradiant phase transition (in the conventional sense) can exist for very specific geometries, but must be understood as a ferroelectric instability~\cite{Keeling2007,Griesser2016,Emelanov1976}, which is essentially unaffected by the coupling to the resonator mode. Nevertheless, this transition is still associated with a characteristic kink in the vacuum fluctuations of the gauge-invariant voltage and flux degrees of freedom. In the non-perturbative regime significant corrections from this classical picture arise due to a hybridization of individual dipoles and photons. Most importantly, in this regime the cavity induces a collective anti-ferroelectric interaction, which favors subradiant ground states where the dipoles tend to anti-align and decouple from the field mode~\cite{Jaako2016}. In this regime also a new transition between superradiant and subradiant ground states becomes possible. These preliminary findings already show that for very strong interactions the physics of cavity QED can differ significantly from the usual picture conveyed by discussions of Dicke or Hopfield-type~\cite{Hopfield} models and that many surprising aspects of USC physics are still unexplored.

The remainder of the paper is structured as follows. After introducing in Sec.~\ref{sec:Model} the setup and the quantities of interest, we first discuss in Sec.~\ref{sec:Classical} the polaritonic eigenmodes and instabilities of classical systems of dipoles in a cavity. In Sec.~\ref{sec:Hamiltonian} we then derive a minimal quantum mechanical model for this system, which after some further simplifications is used in Sec.~\ref{sec:VacuumStates} to investigate the different ground states of cavity QED. We conclude our work in Sec.~\ref{sec:Conclusions} by connecting the findings of this work to different experimental platforms.

\section{Cavity QED: A toy model}\label{sec:Model}

\begin{figure}[t]
  \centering
    \includegraphics[width=\columnwidth]{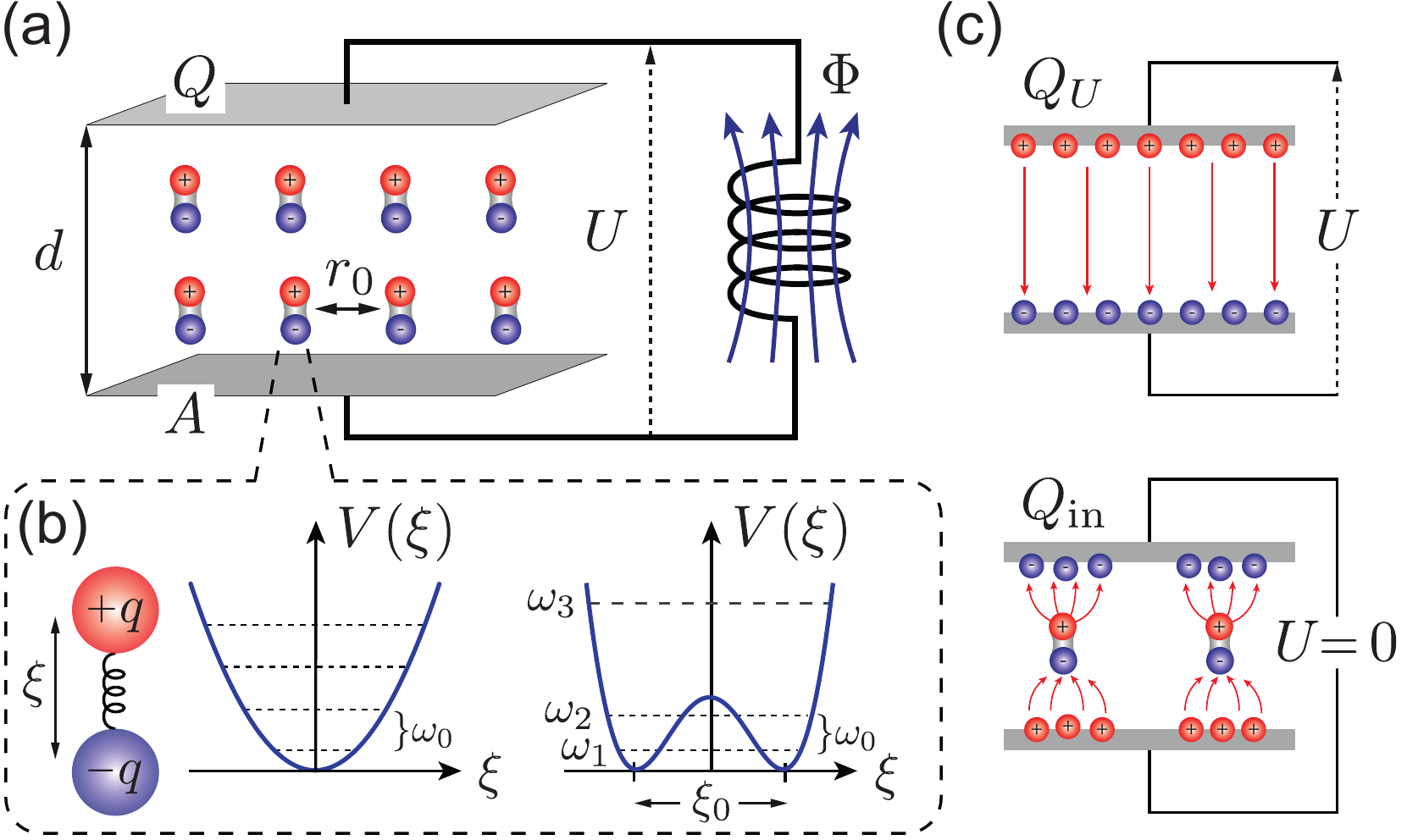}
      \caption{(a) Sketch of the cavity QED setup considered in this work. (b) Different effective potentials $V(\xi)$ for the dipole variable $\xi$ are used to model either harmonic or two-level dipoles of frequency $\omega_0$.  (c) Illustration of the two  different contribution to the total charge $Q=Q_U+Q_{\rm in}$ on the upper capacitor plate. See text for more details.}
      \label{Fig1Setup}
\end{figure}

We consider a setup as shown in Fig.~\ref{Fig1Setup}(a), where $N$ dipoles are coupled to the electric field of a lumped-element $LC$ resonator. The resonator has a bare resonance frequency $\omega_c=\sqrt{1/LC}$, where $C$ is the capacitance and $L$ the inductance of the circuit. This frequency is far separated from all higher order electromagnetic resonances such that the $LC$ resonator is well-described by a single harmonic oscillator mode. The dipoles are assumed to be fixed at positions $\vec r_i=(x_i,y_i,z_i)$ and formed by a pair of charges $+q$ and $-q$, which are displaced by an amount $\xi_i$ in the direction perpendicular to the plates.
The dynamics of each dipole is  modelled as an (effective) particle of mass $m$ moving in a potential $V(\xi)$, as indicated in Fig.~\ref{Fig1Setup}(b). This allows us to treat both harmonically bound dipoles as well as two-level systems by changing from a quadratic to a double-well potential. For all the following derivations it is assumed that the dipole approximation is valid and that radiative effects as well as magnetic interactions can be neglected. 

The dynamics of the $LC$ resonator is governed by the circuit relations $\dot \Phi= U$ and $\dot Q= - \Phi/L$, where $U$ is the voltage  drop across the capacitor, $Q$ is the total charge on the upper capacitor plate and $\Phi$ is the magnetic flux through the inductor.  In the following we write $Q=Q_U +Q_{\rm in}$, where $Q_U=CU$ is the charge in the absence of the dipoles and  $Q_{\rm in}= \int_A dxdy \,   \sigma_{\rm in}(x,y)$ is the total charge induced by the dipoles when $U=0$ [cf. Fig.~\ref{Fig1Setup}(c)]. The induced surface charge density, $\sigma_{\rm in}(x,y)$, depends on the exact distribution of dipoles and can vary strongly across the capacitor plate of total area $A$.  With these definitions we obtain the equation of motion for the flux variable $\Phi$,
\begin{equation}\label{eq:EOMCircuit}
 C \ddot \Phi +\frac{\Phi}{L} =  -\dot Q_{\rm in} \simeq \frac{q}{d}\sum_i \dot \xi_i.
\end{equation}
In the last step we have used the fact that sufficiently far away from the edges of the capacitor the total surface charge induced by a single dipole is $-q \xi_i/d$, where $d$ is the distance between the plates (see App.~\ref{app:DipoleDipole}).  This approximation is not essential, but  results in a convenient  homogeneous dipole-resonator interaction. 

Based on the assumptions stated above, the equations of motion for the  dipole variables $\xi_i$ are $m \ddot \xi_i + V_i'(\xi_i) =   q E(\vec r_i)$, where $E(\vec r_i)$ is the total  electric field at the position of the $i$-th dipole. We decompose this field into two parts, 
\begin{equation}
qE(\vec r_i)= -\frac{q}{d} \dot \Phi  - m \omega^2_{\rm p}  \sum_{i,j}   \mathcal{D}_{ij} \xi_j,   
\end{equation}
where we introduced the plasma frequency, 
\begin{equation}
\omega_{\rm p}^2 = \frac{q^2 }{\varepsilon_0 m r_0^3},
\end{equation}
as the characteristic frequency scale related to the interaction between two neighboring dipoles separated by a distance $r_0$. The dimensionless coupling parameters $\mathcal{D}_{ij}\sim \mathcal{O}(1)$ account for the exact spatial dependence of dipole-dipole interactions. In free space we would simply obtain
\begin{equation}\label{eq:Dij}
\mathcal{D}_{ij}= \frac{r_0^3}{4\pi }    \frac{ |\vec r_{ij}|^2 - 3 (\vec r_{ij}\cdot \vec e_z)^2 }{|\vec r_{ij}|^5},
\end{equation}
where $\vec r_{ij}=\vec r_i-\vec r_j$, but  the capacitor plates can strongly modify this dependence due to the presence of additional image charges \cite{Perram1996}. The numerical evaluation of the $\mathcal{D}_{ij}$ in this confined geometry is detailed in App.~\ref{app:DipoleDipole}. Note that each dipole also interacts with its own image charges and $\mathcal{D}_{ii}\neq0$. In the following we absorb this self-interaction into a redefinition of the potential, i.e., $V_i(\xi_i)+m\omega_{\rm p}^2\mathcal{D}_{ii}\xi_i^2/2\rightarrow V'_i(\xi_i)\simeq V(\xi_i)$, which, for the sake of simplicity, is assumed to be approximately the same for all dipoles. 

In summary, we obtain the following equations of motion for the dynamical variables $\xi_i$,
\begin{equation}\label{eq:EOMDipoles}
m \ddot \xi_i +  V'(\xi_i)  +    m \omega_{\rm p}^2 \sum_{j\neq i}  \mathcal{D}_{ij}  \xi_j   =  -\frac{q}{d}\dot \Phi.
\end{equation}
Together with Eq.~\eqref{eq:EOMCircuit}, this set of equations specifies a minimal model for a cavity QED system consisting of multiple electric dipoles coupled to a single electromagnetic mode.

\section{Polaritons, instabilities and geometry}\label{sec:Classical}
Before we proceed with the quantization of our model, it is instructive to consider first a few basic properties of this system in the limit of a large number of harmonically bound dipoles, i.e., $V(\xi)=m\omega_0^2 \xi^2/2$. For a sufficiently homogeneous system the cavity will couple primarily to the collective variable $\mathcal{Z}=\sum_i \xi_i/\sqrt{N}$, where all dipoles oscillate in phase. By ignoring for now the weak admixing of other excitation modes due to dipole-dipole interactions,  we arrive at a reduced set of two coupled oscillator equations 
\begin{eqnarray}\label{eq:EQM_Z}
\ddot {\mathcal{Z}} +   \left(\omega_0^2+\eta \omega_{\rm p}^2\right)  \mathcal{Z}   &=& -  \frac{q}{dm } \sqrt{N}\dot \Phi , \\
   \ddot \Phi +\omega_c^2 \Phi &=&    \frac{qd }{V\varepsilon_0} \sqrt{N}  \dot {\mathcal{Z}}. \label{eq:EQM_Phi}
\end{eqnarray}
Here we have assumed a parallel plate capacitor with volume $V=Ad$ and capacitance  $C=\varepsilon_0 A/d$ and introduced the dimensionless parameter  
\begin{equation}\label{eq:eta}
\eta = \frac{1}{N}  \sum_{i\neq j}   \mathcal{D}_{ij}.
\end{equation}
This geometrical constant captures the average influence of dipole-dipole interactions in a homogeneously polarized sample and is closely related (but not identical) to the usual depolarization factor of dielectric bodies~\cite{LandauECM}.  Its value depends on the lattice configuration, the shape of the dipole ensemble, and the metallic boundaries, but for a fixed minimal distance $r_0$, it does not scale with the number of dipoles.

\subsection{Polaritons and instabilities}
From Eqs.~\eqref{eq:EQM_Z} and \eqref{eq:EQM_Phi} we readily obtain two polaritonic eigenmodes  
with frequencies (see App.~\ref{app:Polaritons})
\begin{equation}\label{eq:OmegaPM}
\Omega_\pm^2 = \frac{ \omega_{d}^2 + \omega_c^2 +\nu\omega_{\rm p}^2 \pm \sqrt{  (\omega_{d}^2 + \omega_c^2 +\nu \omega_{\rm p}^2)^2-4 \omega_{d}^2\omega_c^2 }}{2},
\end{equation}
where $\omega_{d}=\sqrt{\omega_0^2+\eta \omega_{ \rm p}^2}$ denotes the bare oscillation frequency of the interacting ensemble of dipoles. In Eq.~\eqref{eq:OmegaPM} we have used the identity $q^2N/(mCd^2)=q^2N/(\varepsilon_0Vm)  =  \nu \omega_{\rm p}^2$ to express the collective dipole-field coupling in terms of the plasma frequency and the filling factor $\nu = N r_0^3/V$.

\begin{figure}[t]
  \centering
    \includegraphics[width=0.49\textwidth]{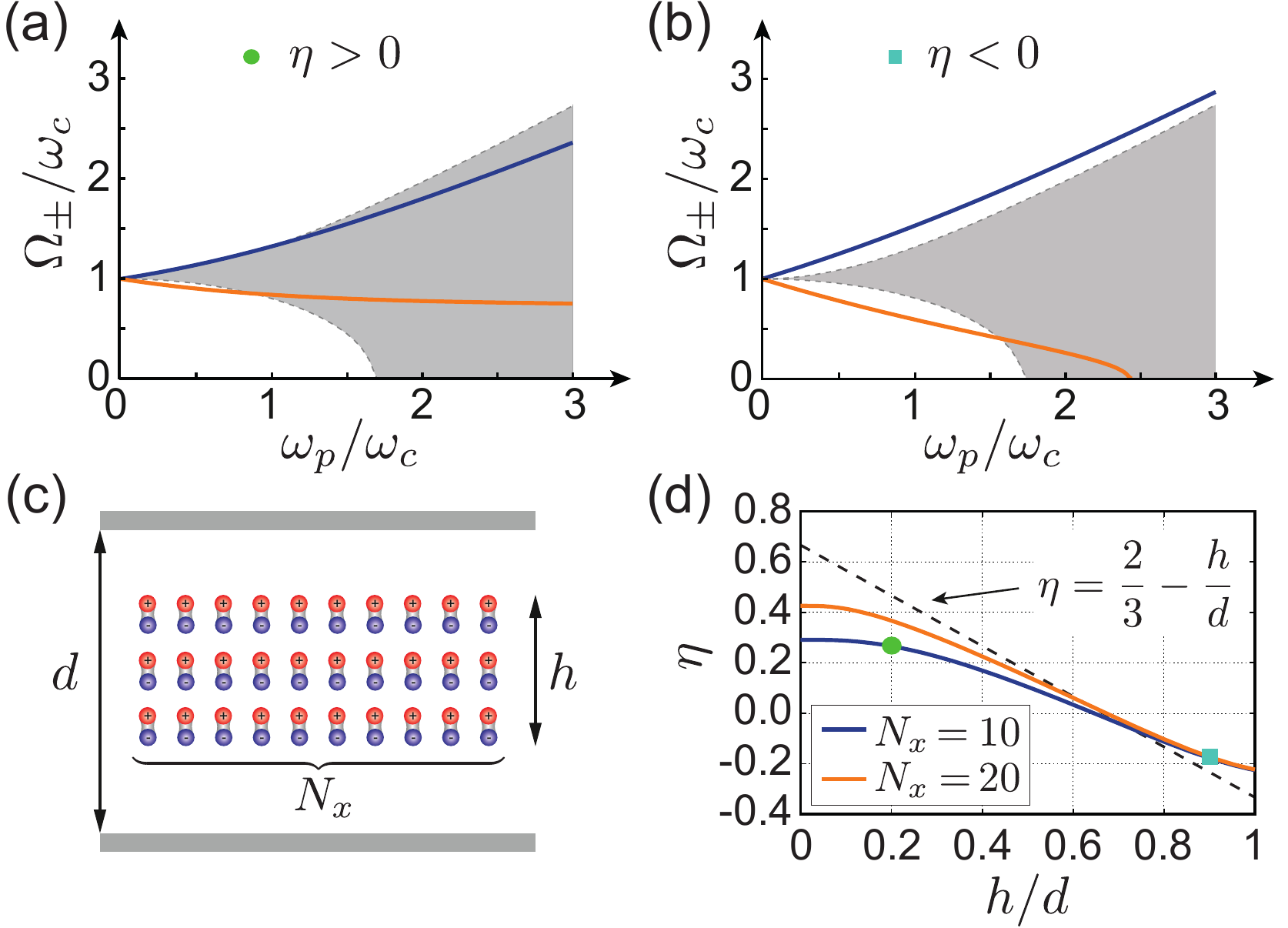}
      \caption{The spectrum of the two bright polariton branches is plotted as a function of $\omega_{\rm p}$ and for $\omega_c = \omega_0$.  In (a) a positive value of $\eta \approx 0.3$ and in (b) a negative value of $\eta \approx -0.2$ has been assumed. In both plots the orange (lower) and the dark blue (upper) lines represent the spectrum obtained from Eq.~\eqref{eq:OmegaPM}, while the shaded area indicates the range of frequencies of all other dark polariton modes obtained from the numerical solution of the full eigenvalue problem (see App.~\ref{app:Polaritons}). (c) Sketch of an ensemble of $N=3\times N_x^2$ dipoles, which are arranged in three layers on a square lattice with spacing $r_0$ and placed between two capacitor plates.  For this configuration the resulting value of $\eta$ is plotted in (d) for varying $d\geq h$ and different $N_x$. The values of $\eta$ and the full coupling matrix $\mathcal{D}_{ij}$ used in the calculations of the polariton spectra in (a) and (b) have been obtained for the case  $N_x=10$ and the values of $ h/d=\nu \approx 0.2$ and $ h/d=\nu \approx 0.9$, respectively.}
      \label{Fig2Classical}
\end{figure}

Figure~\ref{Fig2Classical} shows examples of polaritonic spectra plotted as a function of increasing plasma frequency, i.e., increasing density of dipoles, and for two different values of $\eta$. For resonant interactions, $\omega_0\approx \omega_c$, and for small values of $\omega_{\rm p}$, we observe in both cases the expected normal mode splitting, $\Delta \Omega=\Omega_+-\Omega_-\approx \sqrt{\nu} \omega_{\rm p}$. For larger $\omega_{\rm p}$ and $\eta>0 $ the lower branch approaches a finite value $\Omega_-\approx \omega_c\sqrt{1-\nu/(\nu+ \eta)}$ and remains stable for all parameters.  This behavior is well-known from the study of various solid-state cavity QED systems~\cite{Ciuti2005,Todorov2010,Todorov2012,KenaCohen2013,Geiser2012,Maissen2014,Zhang2016,Hagenmuller2010}, where the regime $\Delta \Omega\sim \omega_c$ is experimentally accessible. In these systems, the observed deviation from a linearly increasing mode splitting is usually derived from the Hopfield model~\cite{Hopfield}, where the $A^2$-term is taken into account. In the opposite case, $\eta<0$, i.e., when dipole-dipole interactions are on average attractive, there exists a critical density or critical plasma frequency, $\omega_{ \rm p}^c=\omega_0/\sqrt{\eta}$, at which the frequency of the lower polariton mode vanishes. Beyond this point the eigenfrequency $\Omega_-$ is imaginary, which means that any excitation of this mode will be exponentially amplified. Therefore, the linear system becomes unstable and a more accurate description of the dipoles must be taken into account.
As shown in more detail in App.~\ref{app:Polaritons}, the critical density at which this instability occurs is determined by a vanishing frequency of the interacting dipole ensemble, i.e. $\omega_{d}=0$, and is independent of the cavity frequency. An experimental signature consistent with such an instability has recently been reported for a 2D hole gas coupled to a THz resonator~\cite{Keller2017}. 

In typical cavity QED experiments the excitation spectrum is inferred from the cavity output field and therefore only the `bright' polariton modes, which are described by Eq.~\eqref{eq:OmegaPM} and involve a large photonic component, are observable. However, there also exist $N-1$ unobservable, i.e., `dark' excitation modes of the dipole ensemble, which due to their spatial profile are almost decoupled from the cavity field (see App.~\ref{app:Polaritons}).  In the examples shown in Fig.~\ref{Fig2Classical}(a) and (b) the frequency range of these additional modes is indicated by the shaded area. We see that even for $ \eta>0$ some of these modes become unstable at high enough densities. Thus, the stability of the experimentally observable bright polarition modes does not necessarily imply the linear stability of the system as a whole.

\subsection{Geometrical considerations}\label{subsec:Geometry}
The shape-dependence of macroscopic thermodynamic properties is a peculiarity of systems interacting via long-range dipole-dipole interactions and is well-known from the study of magnetic or ferroelectric systems. In the limit $N\gg1$ approximate expressions for $ \eta$ can be derived, for example, by treating the dipoles as a continuous medium with polarization density $P(\vec r)$ and solving for the macroscopic electric field  $E_{M}(\vec r)$. The local field $E(\vec r)$ can then be obtained from the relation $E(\vec r)=E_{M}(\vec r)+E_{\rm near}(\vec r)-E_{P}(\vec r)$, where $E_{\rm near}$ is the exact field and $E_{P}=-P/(3\epsilon_0) $ the average field from neighboring dipoles inside a small Lorentz sphere centered around $\vec r$ \cite{LandauECM}. In free space and for dipoles placed on a regular cubic lattice, where $E_{\rm near}\approx 0$, one obtains $ \eta\approx 2/3$ for a disc-shaped ensemble, $\eta\approx 0$ for a spherical ensemble and $\eta \approx -1/3$ for an elongated, cigar-like configuration \cite{LandauECM}. Importantly, these values are modified in the presence of the capacitor \cite{Takae2013,Perram1996}, as illustrated in Fig.~\ref{Fig2Classical}(c) and (d) for the case of a flat layer of dipoles placed between two metallic plates. For this geometry we obtain 
\begin{equation}\label{eq:EtaLayer}
\eta \approx \frac{2}{3} -\frac{h}{d},
\end{equation}
where $h$ is the thickness of the dipole layer. Therefore, the presence of the metallic boundaries can have a substantial effect and bring the system from a stable to an unstable configuration as the distance between the plates is decreased \cite{Bratkovsky2008, Levanyuk2016}. For a single layer of dipoles placed on a triangular lattice and $d\gg r_0$ we obtain $\eta\approx 0.88$ \cite{Nijboer1958,Dell'Anna2016}, while 
the minimal possible value of $\eta=- {\rm Zeta}(3)/\pi\approx -0.38$ is obtained for a line of dipoles placed on top of each other.   

In view of Eq.~\eqref{eq:EtaLayer} it is important to keep in mind that our definition of the potential $V(\xi)$  includes, apart from the external confining potential, also the energy that it takes to separate the charges $+q$ and $-q$. When applying the current analysis to the case of a free electron gas, the limit $\omega_0^2\rightarrow \omega_{\rm p}^2/3$ must be taken to retain this local field contribution. In this limit we recover the usual plasma oscillations, $\omega_{d}\approx \omega_{\rm p}$, for $\nu\rightarrow 0$ and $\omega_{d}\rightarrow 0^+$ for $\nu\rightarrow 1$.

\subsection{Discussion}

From the basics properties of polaritonic systems discussed in this section we can already make the following important observations. (i) Both the collective dipole-field coupling, $\Delta \Omega \sim \omega_{\rm p}$, as well as the strength of direct dipole-dipole interactions, $\sim \omega_{\rm p}^2$, scale with the density and cannot be treated as independent effects. In particular, in the USC regime, where $\omega_{\rm p}\sim \omega_c,\omega_0$, the effect of dipole-dipole interactions plays a dominant role and must be fully taken into account. (ii) A cavity QED system of dipoles coupled to a single electromagnetic mode can exhibit an instability. This instability is induced by dipole-dipole interactions and therefore depends on details like the shape of the ensemble or the lattice configuration. This explains, why many  no-go- and counter-no-go-theorems for superradiant phase transitions, which either completely omit dipole-dipole interactions or do not treat them in all detail, come to very different conclusions. (iii) Most importantly, if an instability exists, it is \emph{solely} induced by dipole-dipole interactions and not influenced by the frequency or other properties of the resonator mode.  This observation contradicts the usual picture conveyed by discussions of the Dicke model, where the transition into the superradiant phase is commonly misinterpreted as being induced by the coupling to a dynamical field mode. Of course, adding the metallic plates in the first place can still substantially modify the properties of the confined system of dipoles compared to its counterpart in free space.

\section{Cavity QED Hamiltonian}\label{sec:Hamiltonian}
Our goal is now to derive a minimal quantum mechanical model for the cavity QED system described in Sec.~\ref{sec:Model}, which is also applicable for highly non-linear dipolar systems and for arbitrary coupling strengths. As a starting point for this derivation we consider the Lagrangian $\mathcal{L}\equiv \mathcal{L}(\Phi,\dot\Phi,\{\xi_i\},\{\dot \xi_i\})$ of the form
\begin{equation}\label{eq:Lagranigan}
\begin{split}
\mathcal{L}= & C \frac{\dot \Phi^2}{2} - \frac{\Phi^2}{2L}  +  \dot\Phi Q_{\rm in}   \\
&+  \sum_i \left[\frac{m }{2} \dot \xi_i^2- V(\xi_i)\right]-\frac{m  \omega_{\rm p}^2}{2} \sum_{i,j}  \mathcal{D}_{i \neq j}  \xi_i  \xi_j,
\end{split}
\end{equation}
from which the equations of motion~\eqref{eq:EOMCircuit} and~\eqref{eq:EOMDipoles} can be derived. 
For this Lagrangian, the resulting canonical momenta are 
\begin{equation}
\Pi =\frac{\partial \mathcal{L}}{\partial \dot\Phi} = C \dot \Phi + Q_{\rm in}, \qquad  \pi_i =\frac{\partial \mathcal{L}}{\partial \dot \xi_i}= m \dot \xi_i =p_i,
\end{equation}
and correspond to the total charge $Q$ on the capacitor plate and the kinetic momenta, respectively. By following the usual  quantization procedure we obtain the Hamilton operator 
\begin{equation}\label{eq:Hfull}
H=  \frac{(Q-Q_{\rm in})^2}{2C} + \frac{\Phi^2}{2L}  + \sum_i H_{d}^i + \frac{m  \omega_{\rm p}^2}{2} \sum_{i,j}  \mathcal{D}_{i \neq j}  \xi_i  \xi_j,
\end{equation}
where $ H_{d}^i=p_i^2/(2m)+  V(\xi_i)$ and $\Phi$, $Q$, $\xi_i$ and $p_i$ are now operators obeying the commutation relations $[\Phi,Q]=[\xi_i,p_j]=i\hbar\delta_{ij}$.  Using $Q_{\rm in}\simeq -q\sum_i\xi_i/d$, Hamiltonian~\eqref{eq:Hfull} can further be expanded  in terms of the operators $\xi_i$, 
\begin{equation}\label{eq:HExpanded}
\begin{split}
H=  &\frac{Q^2}{2C} + \frac{\Phi^2}{2L}   + \frac{q}{C d} Q \sum_i \xi_i   + \frac{q^2}{2Cd^2}\sum_{i,j} \xi_i \xi_j  \\
&   +\sum_i \left[ \frac{p_i^2}{2m}+  V(\xi_i)\right] + \frac{m  \omega_{\rm p}^2}{2}  \sum_{i\neq j}  \mathcal{D}_{ij} \xi_i \xi_j.
\end{split}
\end{equation}
This result represents the full Hamilton operator for the model cavity QED system considered in this work. 

Equation~\eqref{eq:HExpanded} shows that apart from the expected collective coupling of all dipoles to the `charge' of the $LC$ resonator, there are two additional dipole-dipole interaction terms $\sim \xi_i\xi_j$. Since, by construction of our model, the term $\sim \mathcal{D}_{ij}$ already accounts for all direct interactions between the dipoles, the additional presence of the last term in the first line of Eq.~\eqref{eq:HExpanded} 
is very counterintuitive. However, as can be seen from Eq.~\eqref{eq:Hfull}, this term simply arises from expressing the electrostatic energy contribution, $CU^2/2$, in terms of the canonical charge $Q$. It is thus  merely an artifact of our choice of variables and should not be associated with a physical interaction. The inclusion of this term is nevertheless  crucial to recover  the correct equations of motion~\eqref{eq:EOMDipoles} from the relation $m\ddot \xi_i= i/\hbar [H,p_i]$. This subtle difference between  apparent interaction terms in the Hamiltonian and real physical couplings is a common source of confusion in the interpretation of cavity QED models.

\subsection{The $P^2$-term}
In our model, the distribution of point-like dipoles corresponds to a polarization density
\begin{equation}\label{eq:PolarizationDensity}
\vec P(\vec r)= q  \vec e_z  \sum_i  \xi_i \delta (\vec r_i -\vec r).
\end{equation}
For  a sufficiently dense and homogeneous ensemble, where $ \vec P(\vec r)\simeq  \vec e_z P$, we can identify 
$Q_{\rm in}\epsilon_0/(Cd)\simeq -P$ with the polarization density and $Q\epsilon_0/(Cd) \simeq \epsilon_0U/d - P=-D$ with the displacement field $\vec D(\vec r)\simeq \vec e_z D$. With these identifications, Hamiltonian~\eqref{eq:HExpanded} can be directly related to the Hopfield model expressed in the electric dipole gauge \cite{Todorov2012,Bamba2014,PhotonsAndAtoms,ScheelBook},
\begin{equation}\label{eq:HamHopfield}
\begin{split}
H_{\rm HM}=& H_{\rm matter}+ \int d^3 r \,\frac{[\vec D(\vec r)-\vec P(\vec r)]^2}{2\epsilon_0} + \frac{\vec B^2(\vec r)}{2\mu_0}\\
=&H_{\rm matter}+ \int d^3 r \,\frac{\vec D^2(\vec r)}{2\epsilon_0} + \frac{\vec B^2(\vec r)}{2\mu_0}\\
&-\frac{1}{\epsilon_0} \int d^3 r \,\vec D(\vec r)\cdot \vec P(\vec r) +\frac{1}{2\epsilon_0} \int d^3 r \vec P^2(\vec r).
\end{split}
\end{equation}
Here  $\vec B(\vec r)$ is the magnetic field and $H_{\rm matter}=\sum_i H_{d}^i$ is the Hamiltonian for the matter part.  Therefore, the above-discussed $Q_{\rm in}^2$-contribution plays an equivalent role as the  polarization self-interaction or ``$P^2$-term", which appears in the description of macroscopic polarizable media~\cite{Todorov2010,Todorov2012}. It should be emphasized though that for a discrete polarization density as in Eq.~\eqref{eq:PolarizationDensity},  this polarization self-interaction term results in purely local interactions~\cite{PhotonsAndAtoms,VukicsPRA2012,Vukics2014}
\begin{equation}
\int d^3 r\, \vec P^2(\vec r)  \,\,\longrightarrow \,\, \sum_{i} \xi_i^2.
\end{equation}
The apparent discrepancy between such a local $P^2$-term and the non-local coupling derived in Eq.~\eqref{eq:HExpanded} can be resolved by taking into account that Hamiltonian $H_{\rm HM}$ still contains the coupling of the dipoles to \emph{all} electromagnetic modes. As illustrated in App.~\ref{app:SingleModeHopfield} for a basic geometry,  the coupling to these other  high-frequency modes  introduces effective interactions, which restore the correct non-local  $P^2$- and direct dipole-dipole interaction terms. In other words, starting from the full model $H_{\rm HM}$ in the electric dipole gauge, a single-mode approximation is---independent of the frequency separation---not permitted and various approximate treatments of the $P^2$-term lead to very different physical predictions~\cite{Todorov2012,VukicsPRA2012,Vukics2014,Bamba2014}. Our derivation avoids such complications by including the correct dipole-field and dipole-dipole interactions before passing to a quantum description.

\subsection{Two-level-approximation}\label{subsec:TLA}
Of primary interest in the field of cavity QED is the study of nonlinear quantum phenomena, which arise from the coupling of the harmonic field mode to nonlinear matter, in the simplest case represented by two-level dipoles. In our model we can describe this scenario by considering for each dipole a double-well potential with eigenstates $|\psi_n\rangle$ of energy $\hbar\omega_n$ [cf. Fig.~\ref{Fig1Setup}(b)]. For an appropriate choice of parameters the two lowest tunnel-coupled states $\ket{\downarrow}\equiv|\psi_1\rangle$ and $\ket{\uparrow}\equiv|\psi_2\rangle$ are energetically well-separated from all higher excited states and the dynamics of the dipoles can be restricted to this two-level subspace. Under such conditions we can approximate 
\begin{equation}\label{eq:TLSApprox}
H_{d}^i 
 \approx  \frac{\hbar \omega_0}{2}\sigma_z^i,\qquad \xi_i \approx  \frac{\xi_0}{2}\sigma_x^i,
\end{equation}
where the $\sigma_k$ are the usual Pauli operators, $\omega_0=\omega_2-\omega_1$ is the  transition frequency between the two lowest states and $\xi_0=2 \langle \downarrow \! | \xi_i|\!\uparrow\rangle$ is the separation between the wells. According to Eq.~\eqref{eq:TLSApprox}, the  definition of $\omega_0$  does not include a small renormalization of the potential from the additional term $\sim \xi_i^2$ in Eq.~\eqref{eq:HExpanded}. This approximation is justified when $\nu m\omega_{\rm p}^2\xi_0^2/N \ll \hbar |\omega_3-\omega_2|$, which can be achieved for a sufficiently nonlinear potential. Note, however, that for weakly nonlinear systems, for example, superconducting transmon qubits, this renormalization term is highly relevant and constraints the resulting coupling constant to $g<\sqrt{\omega_c\omega_0}$~\cite{Jaako2016,Bosman2017}. 

Within the validity of the two-level approximation and by expressing the resonator variables in terms of annihilation and creation operators, 
$\Phi  = -i \sqrt{\hbar/(2C \omega_c)}(a^\dag-a)$ and  $Q  =  \sqrt{\hbar C\omega_c/2}(a^\dag+a)$, 
we finally obtain the cavity QED Hamiltonian
\begin{equation}\label{eq:HcQED}
\begin{split}
H_{\rm cQED}= &\,\,\hbar \omega_c a^\dag a + \frac{\hbar g}{2} (a+a^\dag)\sum_i \sigma_x^i +\frac{\hbar \omega_0}{2}  \sum_i  \sigma^i_z  \\
  & +  \frac{\hbar g^2}{4\omega_c} \sum_{i,j}\left(1 + \frac{N}{\nu}\mathcal{D}_{ij}\right) \sigma_x^i\sigma_x^j.  
\end{split}
\end{equation}
In this expression we have adopted a notation more familiar in the field of quantum optics and introduced the single-dipole coupling constant  
\begin{equation}\label{eq:g}
g= \frac{q\xi_0}{Cd\hbar }\sqrt{\frac{\hbar C\omega_c }{2}}= \sqrt{\omega_c \frac{m \omega_{\rm p}^2\xi_0^2}{2\hbar}\frac{\nu}{N}}.
\end{equation}
For weak couplings, $g\rightarrow 0$, the second line in Eq.~\eqref{eq:HcQED} can be neglected and $H_{\rm cQED}$  reduces to the standard Dicke model with a collective coupling constant $G=g\sqrt{N}$. When this coupling becomes comparable to $\omega_c$, the Dicke model is no longer valid and the effect of dipole-dipole interactions and the $P^2$-term must be taken into account. Note that both contributions scale as $\sim G^2/\omega_c$, as will become more apparent in the discussion below.


\subsection{Coupling parameter}\label{subsec:Coupling}
In Eq.~\eqref{eq:g} we have related the coupling constant $g$ to the cavity frequency $\omega_c$ and the plasma frequency $\omega_{\rm p}$ such that for a harmonic dipole, where $\xi_0/2=\xi_{\rm HO}=\sqrt{\hbar/(2m\omega_0)}$, the results of Sec.~\ref{sec:Classical} are recovered, i.e., $\Delta\Omega=G\approx \sqrt{\nu}\omega_{p}$, when $\omega_0 \approx \omega_c$. 
In the few dipole, quantum regime the key quantity of interest is the ratio $g/\omega_c=\sqrt{2\pi \alpha}$, which can be expressed in terms of the dimensionless parameter (see also Ref.~\cite{Devoret2007})
 \begin{equation}\label{eq:alpha}
\alpha  = \alpha_{\rm fs} \left(\frac{\xi_0}{d}\right)^2 \left(\frac{q}{e}\right)^2 \frac{Z}{Z_0}.
\end{equation}
Here $e\simeq 1.6\times 10^{-19}$ C is the elementary charge, $Z=\sqrt{L/C}$ the circuit impedance, and $Z_0=\mu_0/\varepsilon_0\approx 377\,\Omega$ the impedance of free space. For an electromagnetic mode with $Z\approx Z_0$ and elementary dipoles of charge  $q=e$ the maximal value of $\alpha$ is set by the finestructure constant $\alpha_{\rm fs}$, which is reached when the size of a dipole is comparable to the size of the cavity, $\sim d$. This illustrates the natural bound on the coupling parameter stated in Eq.~\eqref{eq:Bound}, which can also be obtained for an optical mode confined to a volume $V\approx \xi_0^3$, electric transitions between Landau levels~\cite{Hagenmuller2010}, etc. Eq.~\eqref{eq:alpha} shows that this bound can be reached or even overcome by using artificial atoms like superconducting qubits or quantum dots coupled to tailored circuit resonances with $Z\gg Z_0$ \cite{Bosman2017,Stockklauser2017}. In view of Eq.~\eqref{eq:Bound}, one can then reinterpret such artificial setups as regular cavity QED systems with an effective finestructure constant $\alpha\sim\mathcal{O}(1)$. This analogy establishes an interesting connection to the underlying theory of QED and motivates the study of cavity QED systems in the regime $\alpha\gtrsim 1$ ($g/\omega_c\gtrsim 2.5$), where the electromagnetic interaction can no longer be considered as weak.

Another interesting and in practice  useful observation is that the coupling strength is bounded by 
\begin{equation}
\frac{g}{\omega_c} \leq \frac{q}{2Q_0},
\end{equation}
where $Q_0=\sqrt{\hbar C\omega_c/2}$ is the magnitude of the zero-point charge fluctuations. 
The non-perturbative regime is thus equivalent to the condition that the charge induced by single dipole exceeds the quantum fluctuations of the charge on the capacitor plate. Note that similar bounds can also be obtained for other cavity QED implementations. For example, for a flux qubit coupled inductively to a microwave cavity the coupling is bounded by the ratio $g/\omega_c \leq \Phi_q/(2\Phi_0)$, where $\Phi_q$ is the flux of a qubit state and $\Phi_0=\sqrt{\hbar/(2C \omega_c)}$ the magnitude of the zero-point flux fluctuations of the cavity.

\subsection{Gauge non-invariance}
As discussed  above,  Hamiltonian $H_{\rm cQED}$ represents a cavity QED model in the electric  dipole gauge, which is derived from the Lagrangian $\mathcal{L}$ in Eq.~\eqref{eq:Lagranigan}. The quantization of the electromagnetic field in free space is  commonly performed in the Coulomb gauge, where the so-called minimal coupling Hamiltonian emerges as the fundamental model for light-matter interactions~\cite{PhotonsAndAtoms}. In the current setup, the Coulomb gauge is represented by the Lagrangian,
\begin{equation}\label{eq:GaugeTrans}
\mathcal{L}_{C} =\mathcal{L} - \frac{d}{dt}(\Phi Q_{\rm in}),
\end{equation}
which is related to $\mathcal{L}$ by a canonical transformation.  In this gauge the canonical momenta are 
\begin{equation}
\Pi=\frac{\partial \mathcal{L}}{\partial \dot\Phi} = C \dot \Phi \equiv Q_{U}, \qquad  \pi_i =\frac{\partial \mathcal{L}}{\partial \dot \xi_i}= p_i +  \frac{q}{d}\Phi.
\end{equation}
The canonical charge is now  proportional to the voltage across the capacitor, while the canonical momenta of the dipoles contain an additional magnetic component. 
The resulting Hamilton operator reads 
\begin{equation}\label{eq:HCoulomb}
\begin{split}
H_{C}=  &\frac{Q_{U}^2}{2C} + \frac{\Phi^2}{2L}  + \sum_i \left[ \frac{(p_i-\frac{q}{d}\Phi)^2}{2m}+  V(\xi_i)\right]\\
&+  \frac{m  \omega_{p}^2}{2} \sum_{i\neq j}  \mathcal{D}_{ij}  \xi_i \xi_j,
\end{split}
\end{equation}
and by identifying $\Phi$ with the magnetic vector potential $\vec A$, it can be directly mapped on the minimal coupling Hamiltonian of QED. 
In this representation there are no spurious dipole-dipole interactions, but when expanding the kinetic energy term, we obtain an additional contribution $\sim \Phi^2$.  This is the analogue of the diamagnetic  $A^2$-term and leads to a positive frequency renormalization of the cavity mode. Although Hamiltonian~\eqref{eq:Hfull} and~\eqref{eq:HCoulomb} have a different structure, the canonical transformation in Eq.~\eqref{eq:GaugeTrans} ensures that both Hamiltonians represent the same physical system.   
Indeed, they are related by the unitary transformation $H_{\rm C}= \mathcal{U}  H \mathcal{U}^\dag$, where
\begin{equation}\label{eq:UnitaryGaugeTrans}
\mathcal{U}=e^{i q\sum_i \xi_i \Phi/(d\hbar)} = e^{-i  \Phi Q_{\rm in}/\hbar},
\end{equation}
and for harmonic dipoles it can be explicitly shown that both Hamiltonians reproduce the same spectra \cite{Todorov2012}. 

This gauge equivalence, however, is only guaranteed  when the full Hilbert spaces in each representation are considered. By applying to $H_{C}$ the same two-level approximation as in Sec.~\ref{subsec:TLA}, we obtain an alternative  cavity QED Hamiltonian 
\begin{equation}\label{eq:HcQEDprime}
\begin{split}
H_{\rm cQED}^\prime = \hbar \omega_c a^{\dag} a + \frac{\hbar \omega_{0}}{2} \sum_i \sigma_z^i - i\frac{\hbar g_{C}}{2}(a^{\dag}-a)\sum_i \sigma_y^i
\\
-  \bar{\xi}^{\,2} \frac{\hbar  g_{C}^2 N }{4\omega_{0}}(a^\dag -a)^2 + \frac{ \hbar g^2}{4\omega_{c}} \frac{N}{\nu} \sum_{i,j} \mathcal{D}_{ij}\sigma_x^i  \sigma_x^j,
\end{split}
\end{equation}
where $g_{C}=g\omega_{0}/\omega_c$ and $\bar \xi= 2\xi_{\rm HO}/\xi_{0}\geq1$. The last inequality follows from the Thomas-Reiche-Kuhn sum rule \cite{Lipparini2008}.

By comparing Eq.~\eqref{eq:HcQED} and Eq.~\eqref{eq:HcQEDprime}  it can be readily shown, for example, by setting $\mathcal{D}_{ij}=0$ and $\omega_0=\omega_c$,  that after the two-level approximation the unitary equivalence is lost, i.e., $H_{\rm cQED}\neq \mathcal{U}H_{\rm cQED}^\prime\mathcal{U}^\dag$. While the difference is negligible for weakly coupled systems, the obvious question arises: Which is the appropriate model for cavity QED systems in the USC regime? The answer is suggested by the following general relation between the matrix elements of the position and the momentum operator,
\begin{equation}
\langle \psi_n|p|\psi_m\rangle = i m(\omega_n-\omega_m) \langle \psi_n|\xi |\psi_m\rangle.
\end{equation}
This relation shows that for the momentum operator, the coupling to energetically higher states increases with the energy gap. Therefore, for the $(\Phi \sum_ip_i)$-type coupling transitions to states out of the two-level subspace are not systematically suppressed by a large energy denominator and in the Coulomb gauge a two-level approximation is in general not permitted. This basic argument can be verified numerically for explicit examples, which will be detailed elsewhere~\cite{GaugeNonInvariance}.

Note that this gauge non-invariance does not contradict any of the previous models for QED systems with atoms, molecules or intersubband transitions. In these systems $\bar \xi \approx 1$ and the single-dipole coupling $g$ is very weak such that the equivalence between the dipole and the Coulomb gauge still holds. However, once the regime $g/\omega_c\sim 1$ is reached,  the effective cavity QED Hamiltonians derived in different gauges do no longer agree and lead to qualitatively different predictions. From the analysis presented in this section we conclude that $H_{\rm cQED}$ given in Eq.~\eqref{eq:HcQED} represents indeed the correct effective model for two-level dipoles coupled to a single cavity mode, which is valid both in the weak and USC regime.

\section{The vacua of cavity QED}\label{sec:VacuumStates}
Due to the presence of both  short- and long-range dipole-dipole interactions, the properties of Hamiltonian $H_{\rm cQED}$ can be very complex and will also depend in detail on the specific configuration of dipoles. For a qualitative discussion of the possible ground states of cavity QED it is thus preferential to proceed with a further simplification and replace the actual dipole-dipole interactions by the corresponding all-to-all coupling, 
\begin{equation}\label{eq:Replacement}
\frac{N}{4} \sum_{i,j} \mathcal{D}_{ij} \sigma_x^i\sigma_x^j \, \rightarrow \, \eta S_x^2.
\end{equation} 
Here $\eta$ is the dimensionless configuration parameter already defined in Eq.~\eqref{eq:eta} and we have introduced the collective angular momentum operators $S_k=\sum_i\sigma_k^i/2$. This substitution maps the full Hamiltonian $H_{\rm cQED}$ onto the extended Dicke model $(\hbar=1)$,
\begin{equation}\label{eq:EDM}
H_{\rm EDM}= \omega_c a^\dag a  + \omega_0 S_z +  g (a^\dag +a) S_x + \frac{g^2}{\omega_c}\left(1 + \varepsilon\right) S_x^2.
\end{equation}
In this model, dipole-dipole interactions are treated in an averaged way and can  be described by a single parameter $\varepsilon=\eta/\nu$.  The usual Dicke model is recovered as a specific instance of strong ferroelectric couplings, i.e. $\varepsilon=-1$, while the cases of non-interacting ($\varepsilon=0$) or repulsive ($\varepsilon>0$) dipoles 
appear, for example, in the description of intersubband transitions~\cite{Todorov2014} or certain circuit QED settings~\cite{Jaako2016,Armata2017}. Therefore, $H_{\rm EDM}$  interpolates between and extends various other collective cavity-QED Hamiltonians and shows that each of these models can be associated with a different  arrangement of dipoles. We emphasize though that the replacement in Eq.~\eqref{eq:Replacement} is not a systematic approximation and we will discuss some important differences between collective spin models like  $H_{\rm EDM}$ and the full Hamiltonian $H_{\rm cQED}$  in Sec. \ref{subsec:Fullmod} below.

\begin{figure}[t]
  \centering
    \includegraphics[width=\columnwidth]{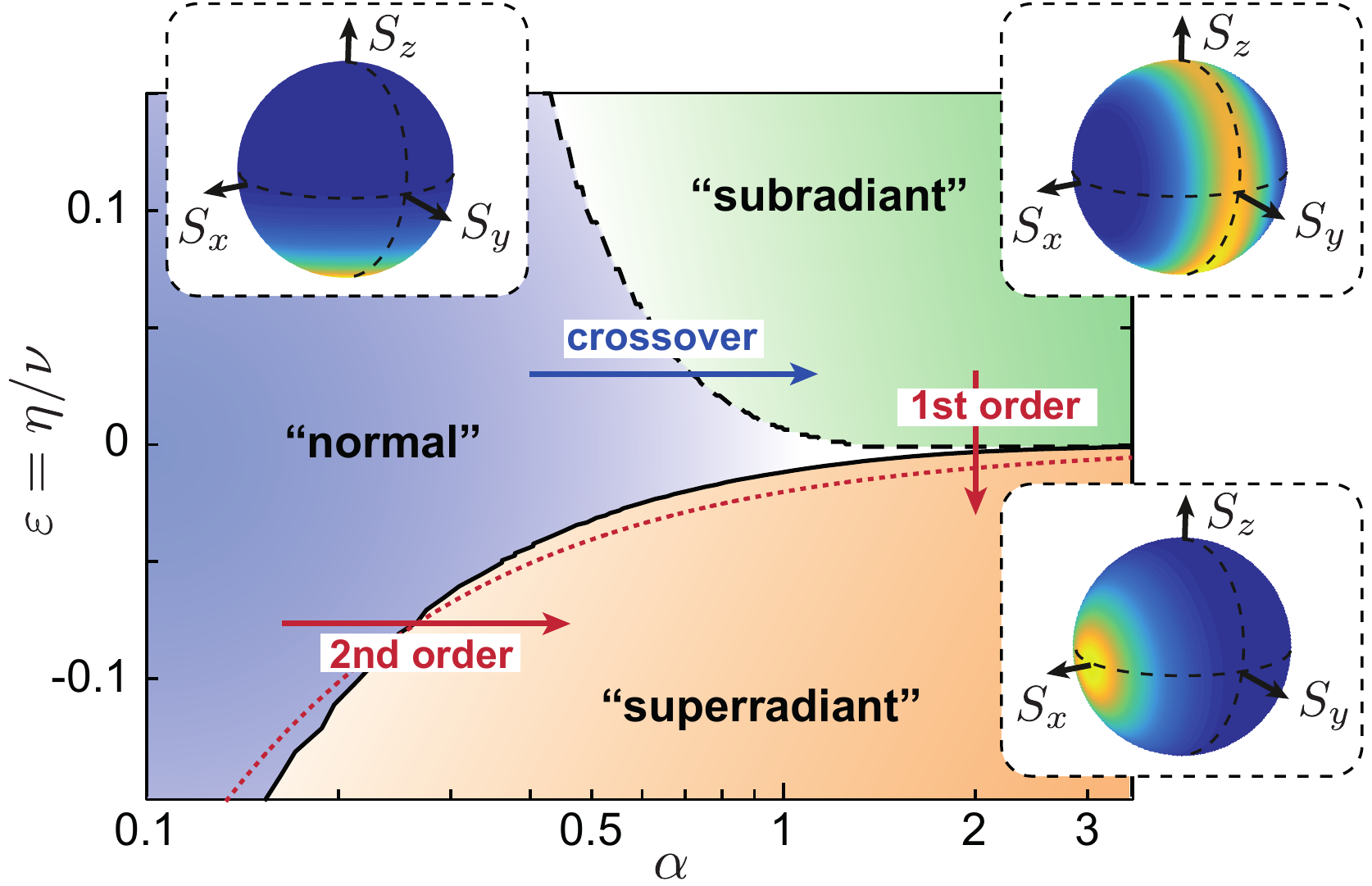}
      \caption{Ground-state phase diagram of the extend Dicke model $H_{\rm EDM}$ as a function of the effective finestructure constant $\alpha=g^2/(2\pi\omega_c^2)$ (horizontal axis) and the average dipole-dipole interaction strength $\varepsilon = \eta/\nu$ (vertical axis).  For this plot $\omega_0/\omega_c=1$ and $N=8$. The red dotted line indicates the value of the critical coupling strength given in Eq.~\eqref{eq:gc} and the other phase boundaries are defined in the text.  For each phase, the insets illustrate the reduced state of the dipoles, $ \rho_d$, in terms of a Bloch-sphere representation. The color shows the value of the Q-function $Q(\vec n)=\langle \vec n|\rho_d|\vec n\rangle\in [0,1]$, where $\vec n$ is a unit vector and $|\vec n\rangle$ the corresponding coherent spin state. Note that for a better visibility, the three insets have been plotted with different colorscales.}
      \label{Fig3Phases}
\end{figure}

Figure~\ref{Fig3Phases} shows a diagram of the ground states of $H_{\rm EDM}$ for different parameters $\alpha$ and $\varepsilon$, which separates into three distinct regimes.  For weak couplings the system is in a \emph{normal} phase, where $\langle a^\dag a\rangle\approx 0$ and $\langle S_z\rangle\approx -N/2$. For increasing $\alpha$ and $\varepsilon<0$ this phase becomes unstable and the system undergoes a transition into a \emph{superradiant} phase. This phase breaks the $Z_2$ symmetry of $H_{\rm EDM}$ and is characterized by a finite expectation value $\langle a\rangle\neq0$ and a finite polarization $\langle S_x\rangle\neq0$.  In the opposite case, $\varepsilon>0$, there is a smooth crossover into a \emph{subradiant} phase. This symmetry-preserving phase is characterized by an anti-aligned spin configuration  with vanishing polarization, $\langle S_z\rangle\approx \langle S_x\rangle\approx 0$, which decouples from the field and therefore $\langle a^\dag a \rangle\approx 0$. For $\alpha>1$ the superradiant and subradiant phase merge and an additional sharp transition between these two phases appears.

\subsection{``Normal phase"}
In the limit $g\rightarrow 0$ the ground state of a cavity QED system is the normal vacuum state with $\langle a^\dag a\rangle=0$ and $\langle S_z\rangle=-N/2$. For finite $g$, corrections to this state can be taken into account by a Holstein-Primakoff approximation~\cite{HolsteinPrimakoff}, where the spins  are replaced by harmonic oscillators, i.e.,  $S_z\rightarrow b^\dag b-N/2$, $S_x\rightarrow \sqrt{N}(b+b^\dag)/2$ and  $[b,b^\dag]=1$. Under this approximation we obtain the quadratic Hamiltonian
\begin{equation}\label{eq:Hlin}
H_{\rm HP} = \omega_c a^\dag a  + \omega_0 b^\dag b + \frac{G}{2}(a+a^\dag)(b + b^\dag) + \frac{D}{4} (b+b^\dag)^2,  
\end{equation}
where $D=\left( 1 + \varepsilon \right)G^2/\omega_c$. $H_{\rm HP}$  can be diagonalized by a Bogoliubov transformation and written in terms of a new set of eigenmode operators $d_\pm$ as $H_{\rm HP}=\Omega_+d^\dag_+d_+ + \Omega_- d_-^\dag d_-$. By identifying $\sqrt{\nu} \omega_{\rm p} \leftrightarrow G\sqrt{\omega_0/\omega_c} $, the eigenfrequencies $\Omega_\pm$ are the same as already obtained for the classical system in Eq.~\eqref{eq:OmegaPM}. This shows that the vacuum state $|G\rangle$ in the normal phase is simply the ground state of the two bright polariton modes described in Sec.~\ref{sec:Classical}. However, one should keep in mind  that $H_{\rm HP}$ does not account for other dark polariton modes, which in the presence of dipole-dipole interaction can lead to important corrections and additional instabilities in the USC regime.

The presence of excitation number non-conserving terms $\sim b^\dag a^\dag$ and $\sim (b^\dag)^2$ in $H_{\rm HP}$ implies that the ground state in the normal phase still exhibits many nontrivial properties when expressed in terms of the original field and matter modes~\cite{Ciuti2006,Fedortchenko2016}. A quantity of interest for the discussion below is the ground state `photon number' $\langle a^\dag a\rangle$, which for moderate couplings is approximately given by
\begin{equation}\label{eq:HPadaga}
\langle a^\dag a\rangle \simeq \frac{Ng^2 \omega_0}{4(\omega_c + \omega_0)^2(\omega_0 +  \varepsilon N g^2/\omega_c )}.
\end{equation} 
Many other ground-state properties of light-matter systems in the linearized regime have been extensively studied in the literature and will not be further elaborated here.

\subsection{``Superradiant phase"}
\begin{figure}[t]
  \centering
    \includegraphics[width=\columnwidth]{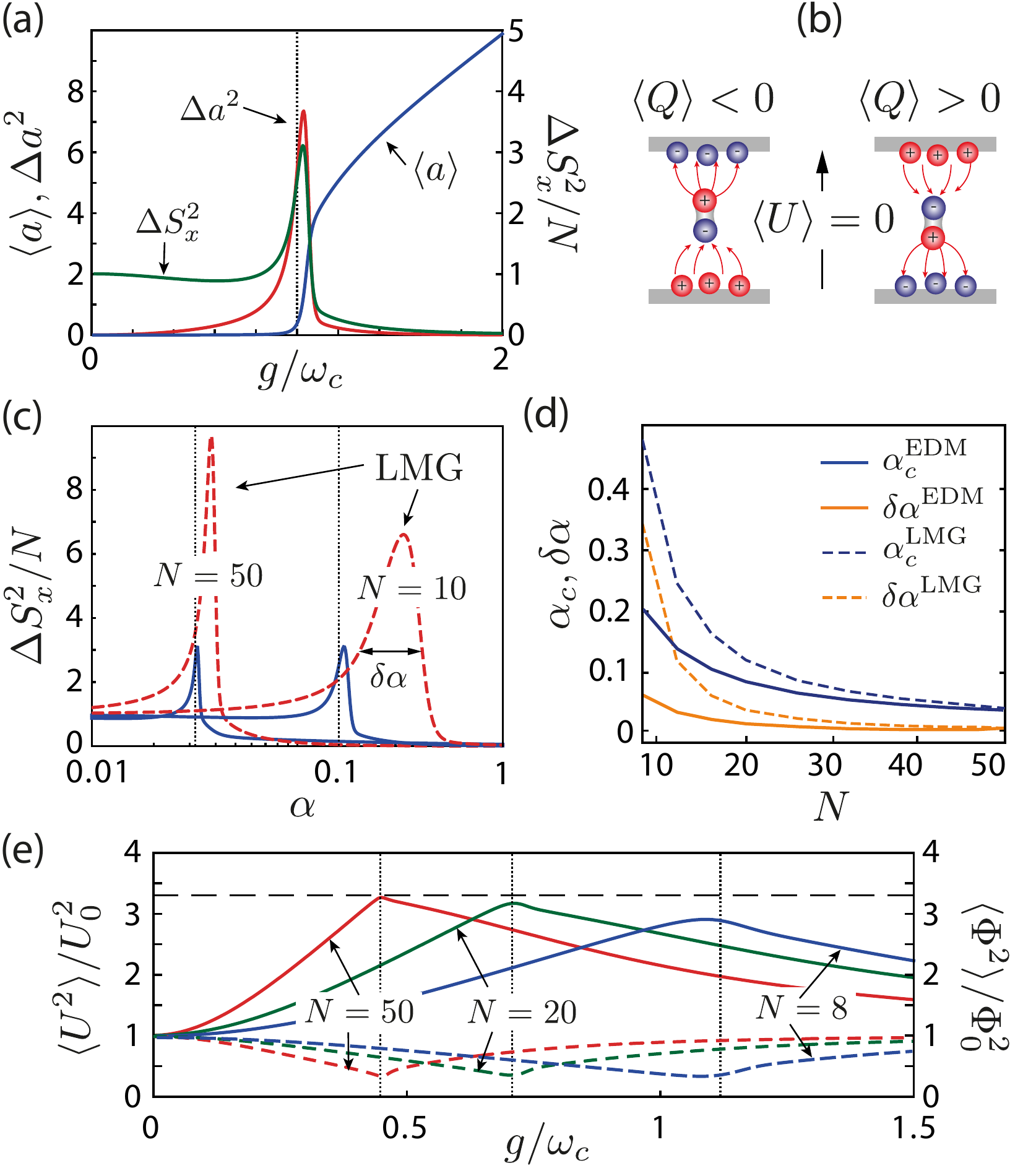}
      \caption{Superradiant phase transition. (a) Dependence of the mean value $\langle a\rangle$ and the spin- and field fluctuations across the superradiant phase transition point. (b) Illustration of the two possible superradiant ground states in terms of polarized dipoles and the corresponding induced charges. 
      (c) Comparison of the spin fluctuations $\Delta S_x$ evaluated with the extended Dicke model and the Lipkin-Meshkov-Glick (LMG) Hamiltonian for two different numbers of dipoles. The value of $\alpha_c$, where the fluctuations reach there maximum, as well as the width of the fluctuations at half of the maximum, $\delta \alpha$ are plotted in (d) for varying $N$. (e) Plot of the fluctuations of the voltage operator $U$ (solid lines) and flux operator $ \Phi $ (dashed lines) for different numbers of dipoles. The horizontal dashed line marks the approximate analytic result given in Eq.~\eqref{eq:U2max}. In all plots $\omega_0=\omega_c$ and a value of $\varepsilon=-0.1$ have been assumed and the vertical dotted lines indicate the analytic phase transition point given in Eq.~\eqref{eq:gc}. In all numerical simulations a symmetry-breaking bias field,  $H_{\rm bias}=\lambda S_x$, where $\lambda/\omega_c=10^{-3}$, has been added to the bare Hamiltonians $H_{\rm EDM}$ and $H_{\rm LMG}$. }
      \label{Fig4SuperR}
\end{figure}

For  increasing coupling $g$ and $\varepsilon <0$ the normal phase eventually becomes unstable and for $N\gg1$ a second order phase transition into a superradiant phase occurs.  This superradiant phase exists for 
\begin{equation}\label{eq:gc}
g \geq g_c=  \sqrt{\frac{\omega_c \omega_0}{-\varepsilon N}},
\end{equation}
and is characterized by a finite polarization of the spins, $\langle S_x\rangle$,  and a finite expectation value of the field mode $\langle a\rangle$. For $g$ close to $g_c$ we obtain \cite{Dimer2007}
\begin{equation}\label{eq:Meanfieldvalues}
\langle a \rangle \simeq \pm \dfrac{Ng}{2\omega_{c}}\sqrt{1 - \left( \dfrac{g_c}{g} \right)^4} ,\quad \langle S_{x}\rangle \simeq \mp \dfrac{N}{2}\sqrt{1 - \left( \dfrac{g_c}{g} \right)^4},
\end{equation}
and $\langle a \rangle \simeq \pm gN/(2\omega_c)$ and $\langle S_{x}\rangle \simeq \mp N/2$ for very large couplings.  As shown in Fig.~\ref{Fig4SuperR}(a), the transition into the superradiant phase is indicated by a sharp peak in the fluctuations of the  polarization, $\Delta S_x^2 =\langle S_x^2\rangle -\langle S_x\rangle^2$, and the field, $\Delta a^2=\langle a^\dag a\rangle-|\langle a\rangle|^2$, and a continuous increase of the order parameter, $\langle a\rangle \sim (g-g_c)^\frac{1}{2}$.  Note that in all our numerical simulations we have added a small symmetry-breaking bias field, which is necessary to deterministically pick one of the two degenerate ground states in the symmetry-broken regime. In Fig.~\ref{Fig3Phases} the maximum of $\Delta S_x^2$ is used to mark the boundary between the normal and the superradiant phase, which even for moderate $N$ agrees reasonably well with the value of $g_c$ obtained from the divergence of $\langle a^\dag a\rangle$ in the Holstein-Primakoff approximation. Thus, the described transition is identical to the conventional superradiant phase transitions discussed within the framework of the Dicke model, but generalized to arbitrary negative values of the interaction parameter $\varepsilon$.

In Eq.~\eqref{eq:gc} the phase boundary between the normal and the superradiant phase is expressed as usual in terms of the dipole-field coupling and the cavity frequency. This form can be very deceiving for identifying the physical origin of this phase transition. By reexpressing $g$ instead  in terms of the original system parameters, Eq.~\eqref{eq:gc} can be rewritten as    
\begin{equation}\label{eq:TransitionPoint}
\hbar \omega_{d} = \hbar \omega_0 + \dfrac{\eta m \omega_{\rm p}^2 \xi_0^2}{2}\leq 0,
\end{equation}
and all cavity-related parameters disappear. While the same is true for the phase transition point of the original Dicke model~\cite{Rzazewski1975,Keeling2007,Vukics2015}, Eq.~\eqref{eq:TransitionPoint} shows that the cancellation of $\omega_c$  is not simply a coincidence. By setting $\eta=0$, the dipole-field coupling can still be arbitrary strong, but no instability occurs. This confirms our observation from above, 
 namely  that the superradiant instability is in essence a ferroelectric instability and not related to the coupling to the dynamical field mode.   

We can further elaborate this point by looking more closely at the physical properties of the superradiant phase. In the current setup a finite expectation value $\langle a\rangle\in \mathbbm{R}$ corresponds to a finite expectation value of the total charge $\langle Q\rangle\sim \langle a+a^\dag\rangle$, which includes charges induced by the dipoles. Therefore, as illustrated in Fig.~\ref{Fig4SuperR}(b), the superradiant ground state simply corresponds to a state of polarized dipoles and the corresponding induced image charge on the capacitor plate. Since the total charge $Q$ is not directly accessible, the more relevant resonator variables to consider are the magnetic flux $\Phi$ and the voltage drop $U$. The latter can be expressed as 
\begin{equation}
U= U_0 \left[ a + a^\dag  + \frac{2g}{\omega_c} S_x\right],
\end{equation}
where $U_0=Q_0/C$. Importantly, the expectation values of the flux and voltage operators are unaffected by the phase transition and we have $\langle \Phi\rangle =\langle U\rangle=0$ in the normal as well as in the superradiant phase. This can  be seen directly from Eq.~\eqref{eq:Meanfieldvalues}, or more generally from the fact that for any stationary state $\langle \dot \Phi\rangle =\langle U\rangle =0$ and $L\langle \dot Q\rangle = - \langle \Phi\rangle =0$. Therefore, although the superradiant phase is conventionally characterized by a finite `field' expectation value $\langle a\rangle\neq 0$, the transition affects the displacement field, $D\sim Q$, and not the electric field, $E\sim U$~\cite{Keeling2007}. On the mean-field level the actual physical properties of the cavity do not change when transitioning between the normal and the superradiant phase. This example illustrates that the imprecise notion of a `photon' annihilation operator $a$ can be very misleading, since depending on the choice of gauge and the setup under consideration this operator  can represent very different physical quantities. 

Given this interpretation in terms a ferroelectric phase transition, where the `radiation mode' does not play a role, it would seem natural to abandon notions of a superradiant transitions and phases all together. It should be kept in mind though that this analogy only concerns average quantities and  strictly holds only in the limit of $N\gg 1$ and $\alpha\ll 1$. When finite-size and strong-coupling effects are taken into account, the presence of the electromagnetic mode can substantially influence  the transition as well as all thermodynamic properties, where excitations on top of the ground state must be taken into account. As an example, we compare in Fig.~\ref{Fig4SuperR}(c) and (d), the predictions from the extended Dicke model with the predictions from the corresponding Lipkin-Meshkov-Glick Hamiltonian~\cite{LMG} 
\begin{equation}
H_{\rm LMG}=  \omega_0 S_z + \varepsilon \frac{g^2}{\omega_c}  S_x^2.
\end{equation}
For $\varepsilon <0$ this Hamiltonian represents a model for ferroelectricity with infinite-range interactions and can  be obtained from $H_{\rm EDM}$ by taking the limit $\nu\rightarrow 0$, but keeping $g^2/\nu$ fixed. We see that for values of $|\varepsilon| N\lesssim 1$, i.e., when the transition already happens at rather large values of $g_c/\omega_c$, the range of fluctuations of $\Delta S_x^2$ as well as the transition point itself are still considerably different. Only for larger numbers, $|\varepsilon| N\gg1$, the two models start to agree better.   Overall we find that the coupling to the cavity mode suppresses fluctuations and generates a sharper transition even for small $N$. This can be in part explained by a dressing of the dipoles with photons, as explained further below. 

Finally, a unique signature of a superradiant transition can be obtained by looking at cavity observables, which do not have a counterpart in ferroelectric models.  As an example, we plot in Fig.~\ref{Fig4SuperR}(e) the behavior of the voltage fluctuations, $\langle U^2\rangle$, across the transition point.  While the gauge non-invariant photon number $\langle a^\dag a \rangle$ diverges at the transition point, the voltage fluctuations remain finite and show a characteristic kink. For $N\gg1$ the position of this kink coincides with the classical transition point $g_c$ and the maximal value of the fluctuations scales approximately as 
\begin{equation}\label{eq:U2max}
\left.\frac{\langle U^2\rangle}{U^2_0}\right|_{g=g_c}\approx \sqrt{1 + \dfrac{1}{|\varepsilon|} \left(\dfrac{\omega_{0}}{\omega_{c}} \right)^2}.
\end{equation}
Interestingly, this maximum does neither scale with $N$ nor the coupling parameter $\alpha$, but the kink vanishes for an interaction dominated system, $\varepsilon\rightarrow \infty$. It thus represents a quantum mechanical signature of a superradiant phase transition, which involves the dynamical cavity mode.  Note that the maximum of $\langle U^2\rangle$ is accompanied a corresponding minimum of the flux fluctuations, $\langle \Phi^2\rangle_{g=g_c}\simeq  \hbar^2/(4C^2 \langle U^2\rangle) $, as expected for a minimum uncertainty squeezed state.

\subsection{``Subradiant phase"}
\begin{figure}[t]
  \centering
    \includegraphics[width=0.48\textwidth]{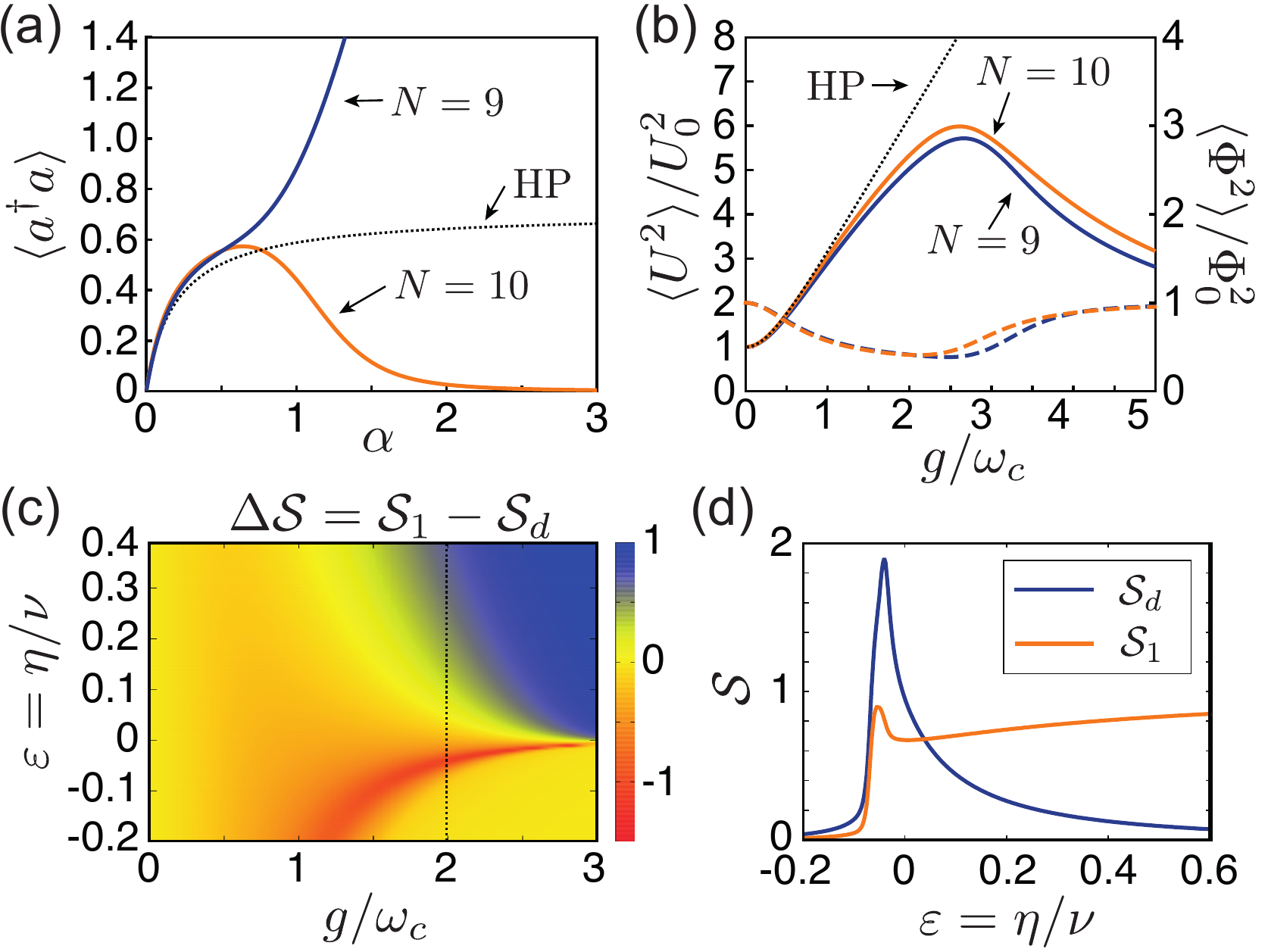}
      \caption{Subradiant phase. 
(a) Plot of the ground state photon number as a function of the coupling parameter $\alpha$ for an even and an odd number of dipoles. (b) Dependence of the voltage (solid lines) and flux (dashed lines) fluctuations on the coupling strength $g$. In both plots the parameters $\varepsilon=0.05$ and $\omega_c = \omega_0 = 1$ have been assumed and the dotted lines shows the corresponding results for $\langle a^\dag a\rangle$ and $\langle U^2\rangle$ obtained from the ground state of the Holstein-Primakoff (HP) Hamiltonian $H_{\rm HP}$ for $N=10$.  
       (c) Plot of the residual single-spin entropy $\Delta \mathcal{S}$ of the ground state of $H_{\rm EDM}$ for $\omega_0=\omega_c$ and $N=4$. (d) The entanglement entropies for a single dipole and for all dipoles are plotted for a fixed $g/\omega_c=2$ and otherwise the same parameters as in (c). In all numerical simulations a symmetry-breaking bias field,  $H_{\rm bias}=\lambda S_x$, where $\lambda/\omega_c=10^{-4}$, has been added to the bare Hamiltonians $H_{\rm EDM}$.}
      \label{Fig5SubR}
\end{figure}

For repulsive dipole-dipole interactions, i.e. $\eta>0$, the linearized Hamiltonian $H_{\rm HP}$ predicts that the normal phase remains stable for arbitrary interaction strengths. For this reason the parameter regime $\eta\geq 0$ and $g\gtrsim \omega_c $ has received little attention in the discussion of USC cavity QED so far. However, although there is indeed no sharp phase transition,  Fig.~\ref{Fig5SubR}(a) clearly shows that the properties of the ground state change significantly when we go beyond the validity of the Holstein-Primakoff approximation into the non-perturbative regime $\alpha\gtrsim 1$. In stark contrast to the superradiant phase, the ground state photon number  in this regime decreases with increasing coupling strength and approaches zero for very large couplings. This behavior has recently been described in the context of circuit QED \cite{Jaako2016} and explained in terms of 
an anti-ferroelectric alignment of the dipoles, which then decouple from the cavity mode. For $N$ even, the resulting ground state is approximately of the form  
\begin{equation}
|G\rangle \simeq |0\rangle \otimes|D_0\rangle,
\end{equation}
where $|D_0\rangle =| s=N/2,m_x=0\rangle$ is the fully symmetric Dicke state with vanishing projection along $S_x$ (see Fig.~\ref{Fig3Phases}), i.e., $S_x|D_0\rangle=0$. For an odd number of dipoles a perfect anti-alignment is not possible and, as shown in Fig. \ref{Fig5SubR}(a), the dipoles and the cavity remain coupled. 


 %

The current analysis and further studies of the full model $H_{\rm cQED}$ below show that the formation of such subradiant ground states is not a peculiarity of superconducting circuits, but rather a general property of non-perturbative cavity QED.  For an even number of dipoles, a possible way to characterize these states is via the decoupling condition
\begin{equation}
\frac{\partial }{\partial g}  \langle a^\dag a\rangle <0,
\end{equation}
which is used in Fig.~\ref{Fig3Phases} to mark the boundary between the normal and the subradiant phase.  It should be pointed out that such a light-matter decoupling can already be predicted within the Holstein-Primakoff approximation, as originally discussed in Ref. \cite{DeLiberato2014} for a multimode cavity QED system. However, for the present single mode scenario, such linear decoupling effects are not observable for the considered parameter range [see, for example, Fig.~\ref{Fig5SubR}(a)]. More specifically,  for the current setting and within the Holstein-Primakoff approximation the ground state photon number, 
\begin{align}
   \left. \lim_{g \rightarrow \infty } \langle a^{\dagger}a \rangle\right|_{\rm HP} = \dfrac{1 + 2\varepsilon - 2\sqrt{\varepsilon(\varepsilon+ 1)}}{4\sqrt{\varepsilon(\varepsilon + 1)}} > 0,
\end{align}
remains finite for very large couplings. Therefore,
the suppression of the photon number below this bound signifies the formation of highly entangled anti-ferroelectric states, which decouple much more efficiently from the cavity than the corresponding squeezed states of the linearized theory.

In Fig.~\ref{Fig5SubR}(b) we  plot the fluctuations of the observable voltage and flux variables and find a very similar qualitative behavior as for the superradiant transition. The voltage fluctuations show again a characteristic peak, which, however, is much smoother and doesn't sharpen  when the number $N$ is increased. Also the position of the maximum doesn't vary significantly as a function of $N$ or $\varepsilon$ and always occurs around $\alpha\approx 1$. The absence of any significant even-odd effects make this peak in $\langle U^2\rangle$ a robust signature for entering the non-perturbative coupling regime. Interestingly, while the subradiant phase is characterized by a strong decoupling of the dipoles from the cavity operator $a$, we find that the level of voltage fluctuations is even higher than in the superradiant phase and also the flux variance $\langle \Phi^2\rangle$ is substantially larger than for a minimal uncertainty state.

Finally, a very interesting property which distinguishes the subradiant from the normal and the superradiant phase, is the high degree of entanglement between the dipoles, while being almost completely disentangled from the cavity mode. This property can be visualized by introducing the two entanglement entropies
\begin{equation}
\mathcal{S}_{1} = -{\rm Tr}\{\rho_1 \log_2(\rho_1)\}, \qquad \mathcal{S}_{d}= -{\rm Tr}\{ \rho_d \log_2(\rho_d) \}.
\end{equation}
Here $\rho_d={\rm Tr}_c \{ |G\rangle\langle G|\}$ is the reduced density operator of the dipoles, and $\rho_1={\rm Tr}_{N-1}\{\rho_d\}$ is the reduced density operator of a single dipole. Therefore, $\mathcal{S}_d$ quantifies the entanglement between the dipoles and the cavity and $\mathcal{S}_1$ the entanglement between a single dipole and the remaining system. In Fig.~\ref{Fig5SubR}(c) and (d) we plot the difference $\Delta \mathcal{S}=\mathcal{S}_{1}-\mathcal{S}_{d}$ and the individual entanglement entropies for different parameter regimes. The plots show that a significant amount of ground-state  entanglement occurs near the superradiant phase transition, but also that this entanglement is established mainly between the cavity and the dipoles. In contrast, when entering the subradiant phase, $\mathcal{S}_d$ is strongly reduced, while the dipoles still remain highly entangled among each other.

\section{Non-perturbative cavity QED}
The analysis in the previous section showed that for most parameter regimes the ground state of $H_{\rm EDM}$ is either a normal vacuum state or a state dictated by strong dipole-dipole interactions.  From the perspective of cavity QED, it is thus most interesting to consider the regime $\eta\approx 0$ and $\alpha\gtrsim 1$, where dipole-dipole interactions play a minor role and the influence of the cavity mode becomes important. As indicated in Fig.~\ref{Fig3Phases}, in this regime the superradiant and subradiant phases approach each other and a new transition between these two very different phases emerges. 

\subsection{Strong-coupling theory}
For the following discussion we return to the full cavity QED Hamiltonian $H_{\rm cQED}$ and focus on the regime $\omega_0\sim\omega_c$ and $\alpha\geq 1$. In this case the coupling to the cavity $\sim g$ and the dipole-dipole interactions $\sim g^2$ dominate over the bare energy splitting of dipoles. It is thus useful to transform into a new basis, which diagonalizes these two terms. This is achieved by a polaron transformation $\tilde H_{\rm cQED}= \mathcal{U} H_{\rm cQED}\mathcal{U}^{\dag}$, where \cite{Jaako2016,Chen2008}
\begin{equation}
\mathcal{U}=e^{ \frac{g}{\omega_c}S_x (a^{\dag} - a)}.
\end{equation}
As a result we obtain
\begin{equation}\label{eq:polaronHam_cQED}
\begin{split}
	\tilde H_{\rm cQED} & = \omega_c a^\dag a + \frac{g^2}{4\omega_c}\frac{N}{\nu} \sum_{i,j}\mathcal{D}_{ij}\sigma_x^i\sigma_x^j
\\
	&+\frac{\omega_0}{2}\left( e^{ \frac{g}{\omega_c}(a^{\dag} - a)} \tilde{S}_{-} + e^{ -\frac{g}{\omega_c}(a^{\dag} - a)} \tilde{S}_{+} \right), 
\end{split}
\end{equation}
where $\tilde{S}_{\pm} = S_z \pm i S_y$ are collective ladder operators with respect to $S_x$.
Note that $\mathcal{U}$ is just the gauge transformation~\eqref{eq:UnitaryGaugeTrans} restricted to the two-level subspace. Therefore, Hamiltonian $\tilde H_{\rm cQED}$ represents the appropriate USC cavity QED Hamiltonian in the Coulomb gauge. By expanding the exponentials in the second line in Eq.~\eqref{eq:polaronHam_cQED} up to first order in $\alpha$, we obtain  
\begin{equation}
	\begin{split}
		\tilde{H}_{\rm cQED} & \approx   \omega_c a^{\dag} a + \omega_{0} S_z - i g\frac{ \omega_0}{\omega_c}(a^{\dag} - a) S_y\\
		&+ \frac{g^2}{2\omega_{c}^2}\omega_0(a^\dag -a)^2 S_z + \frac{ g^2}{4\omega_{c}} \frac{N}{\nu} \sum_{i,j} \mathcal{D}_{ij}\sigma_x^i  \sigma_x^j,
	\end{split}
\end{equation}
which resembles very closely $H_{\rm cQED}^\prime$ given in Eq.~\eqref{eq:HcQEDprime} in the limit of  low excitation numbers. This correspondence is lost when highly excited states or higher-order terms in the coupling parameter are taken into account. 

In the limit $\omega_0\rightarrow 0$, the first line of Eq.~\eqref{eq:polaronHam_cQED} is diagonal in the photon number states $|n\rangle$ and the spin states $|s_i=\pm1\rangle$, where $\sigma_x^i|s_i\rangle=s_i|s_i\rangle$. Therefore, we obtain a set of eigenstates $ |n,\{s_i\}\rangle $ with energies 
\begin{equation}
E^0_{n,\{s_i\}} = n \omega_c  +   \frac{g^2}{4\omega_c}\frac{N}{\nu} \sum_{i,j}\mathcal{D}_{ij}s_is_j.
\end{equation} 
Note that in the original basis the eigenstates represent displaced photon number states,
\begin{equation}\label{eq:DisplacedStates}
|\Psi\rangle_{n,\{s_i\}}= e^{- \frac{g}{\omega_c}S_x (a^{\dag} - a)} |n,\{s_i\}\rangle, 
\end{equation}
with a displacement amplitude $\beta = g/\omega_c \sum_i s_x^i$ proportional to the total spin projection along the $x$ axis. 

For finite $\omega_0$ quantum fluctuations of the dipoles induce finite couplings between different spin projections and different photon number states. For $\alpha\gtrsim 1$ these couplings can be included in second order perturbation theory following Ref.~\cite{Jaako2016}. In the presence of dipole-dipole interactions the result of such a calculation  would still be very involved, since the bare energy levels $E^0_{n,\{s_i\}}$ depend explicitly on the spin configuration.  For the purpose of this work we restrict ourselves to $|\mathcal{D}_{ij}|N/\nu < g^2/\omega_c $, where this dependence can be neglected. By projecting onto the $n=0$ sub-manifold we then obtain the effective spin Hamiltonian
\begin{equation}
H_{S} = \omega_0 e^{-\frac{g^2}{2\omega_c^2}}S_z - \frac{\omega_0^2 \omega_c}{2g^2}\left(\vec{S}^2-S_x^2\right)   + \frac{g^2N}{4\omega_c\nu} \sum_{i,j}\mathcal{D}_{ij}\sigma_x^i\sigma_x^j. 
\end{equation}
From this approximate model we see that the coupling to the cavity mode has two main effects. First, due to the polaronic nature of the eigenstates $|\Psi\rangle_{n,\{s_i\}}$, which contain both dipole and photonic components, the transition frequency $\omega_0$ becomes exponentially suppressed. Second,  virtual excitations of higher photon-number states result in collective dipole-dipole interactions, which  favor states of maximal total spin $S=N/2$, but minimal spin projection along $x$.  

\subsection{Subradiant-to-superradiant phase transition}

Given the effective spin Hamiltonian $H_S$ we can now investigate in more detail the transition between the super- and the subradiant phase, which exist for $\alpha\gtrsim1$. In a first step we will consider again a collective spin model where $\mathcal{D}_{ij}=\eta/N$. In this case the total spin is conserved and we can restrict our analysis to states with $S=N/2$. In terms of the effective finestructure constant we then obtain the LMG model,
\begin{equation}\label{eq:HSColl}
H_{S} = \omega_0 e^{-\pi\alpha}S_z + \left(2\pi \alpha\varepsilon \omega_c + \frac{\omega_0^2 }{4\pi \alpha \omega_c} \right) S_x^2,
\end{equation}
with a renormalized frequency and a modified coupling term. By changing $\varepsilon$ from positive to negative values, the expected transition from the sub- into the superradiant phase can be determined from a Holstein-Primakoff approximation for $H_S$ and we obtain the phase transition point  
\begin{equation} \label{eq:epsilonc}
\omega_0 e^{-\pi\alpha} +  N\left(2\pi \alpha\varepsilon \omega_c + \frac{\omega_0^2 }{4\pi \alpha \omega_c} \right)  = 0.
\end{equation}
By omitting the second term in the brackets we obtain a condition for the critical coupling parameter $\alpha_c$, which is analogous to Eq.~\eqref{eq:gc}, but with a reduced dipole-frequency. This  shows why for small $\varepsilon$ and small $N$ the transition into the superradiant phase occurs at much smaller couplings than predicted by the linearized theory (cf. Fig.~\ref{Fig3Phases}). For $\alpha> 1$ the dipole frequency is fully suppressed and the ground state phase is only determined by the sign of the $S_x^2$-term.  The resulting critical interaction parameter $\varepsilon_c$ is independent of $N$ and given by
\begin{equation} 
\varepsilon_c \simeq -   \frac{\omega_0^2 }{8\pi^2 \alpha^2 \omega_c^2}.
\end{equation}
This result shows that cavity fluctuations stabilize the subradiant phase even beyond $\varepsilon=0$ and that a small, but finite attraction between the dipoles is required to push the system into the superradiant phase. 

\begin{figure}[t]
  \centering
    \includegraphics[width=0.48\textwidth]{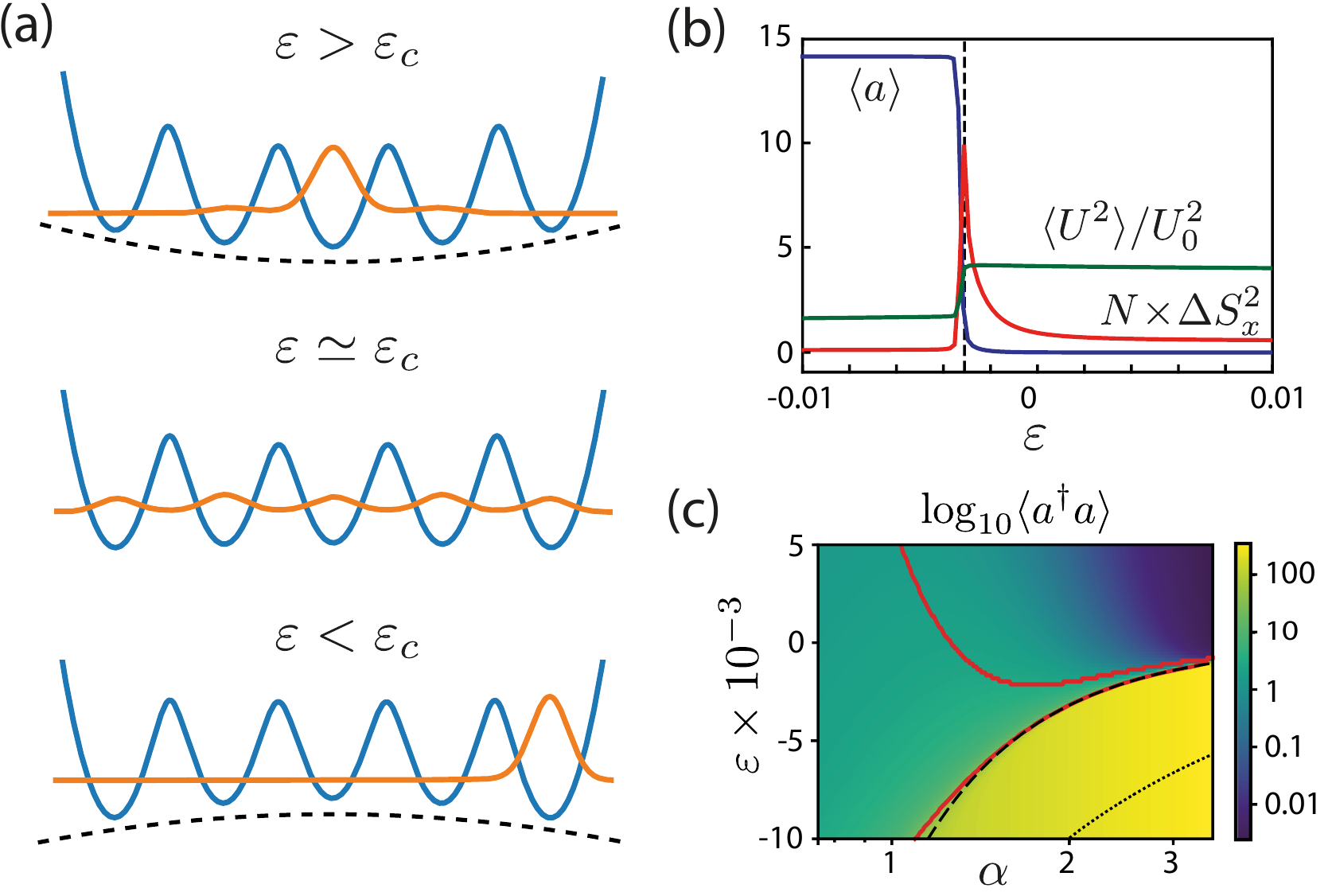}
      \caption{Subradiant-to-superradiant phase transition. (a) The adiabatic potential $ V_{\rm ad}(X) = X^2/2 + E_0(X)$ for the cavity mode is plotted together with the resulting ground state wavefunction for different values of the interaction parameter $\varepsilon$.  (b) Dependence of the mean cavity field and the voltage and spin fluctuations as a function of $\varepsilon$ for $ \alpha = 2 $. (c) Zoom of the ground state phase diagram in the region $ |\varepsilon| \approx 0 $. The color scale shows the ground state photon number and the red solid lines indicate the same phase boundaries as in Fig.~\ref{Fig3Phases}. The dashed line is the critical value $ \varepsilon_c $ given Eq.~\eqref{eq:epsilonc} and the dotted line indicates the value of $ \varepsilon_c $ obtained from the classical transition point in Eq.~\eqref{eq:gc}.  In (a) we have used $N=4$ qubits and in (b) and (c) $N=8$. In all plots $\omega_0=\omega_c$ and a symmetry-breaking bias field,  $H_{\rm bias}=\lambda S_x$, where $\lambda/\omega_c=10^{-3}$, has been assumed.}
      \label{Fig6NonPerturb}
\end{figure}

Figure~\ref{Fig6NonPerturb}(a) illustrates the subradiant-to-superradiant phase transition in terms of the adiabatic Born-Oppenheimer potentials $V_{\rm ad}(X)=X^2/2+E_0(X)$. Here $X=(a+a^\dag)/\sqrt{2}$ is the normalized position quadrature of the cavity mode and $ E_0(X)$ is the ground-state energy of the spin part of the extended Dicke model,
\begin{equation}
H_{\rm EDM} (X)= \omega_0 S_z +  \sqrt{2} g X S_x + \frac{g^2}{\omega_c}\left(1 + \varepsilon\right) S_x^2,
\end{equation}
obtained for different values of $X$. The resulting potentials clearly show the displaced quadratic lobes expected from the displaced oscillator states given in Eq.~\eqref{eq:DisplacedStates} as well as the overall quadratic shift from the $ S_x^2$ term in Eq.~\eqref{eq:HSColl}.  This shift stabilized the subradiant state with $X\approx 0$ for $\varepsilon>\varepsilon_c$, while in the superradiant phase two minima at $X\approx\pm  Ng/\sqrt{2}\omega_c$ emerge. As shown in ~\ref{Fig6NonPerturb}(b) the transition point is indicated by a jump in $\langle a\rangle$ as well as in the voltage fluctuations $\langle U^2\rangle$. This behavior is reminiscent of a first-order phase transition, with the additional peculiarity that at the transition point all the lobes in $V_{\rm ad}(X)$ become energetically degenerate. Finally, Fig.~\ref{Fig6NonPerturb}(c) shows a zoom of the phase diagram in Fig.~\ref{Fig3Phases} with the strongly modified phase boundaries in the non-perturbative regime.

\subsection{Beyond the collective spin approximation}\label{subsec:Fullmod}
In our analysis so far we have primarily focused on the collective spin model where only the averaged dipole-dipole interaction strength appears. For certain cavity QED implementation, in particular in the context of circuit QED, this collective coupling arises naturally from the circuit design~\cite{Jaako2016}, in which case also $H_{\rm EDM}$ becomes exact. However, it is clear that in general the approximation of an arbitrary coupling matrix $\mathcal{D}_{ij}$ by a single parameter $\eta$ can lead to qualitatively very different results. In the following we illustrate the relation between the exact short-range and the collective spin model for two different settings.

\begin{figure}
  \centering
    \includegraphics[width=0.48\textwidth]{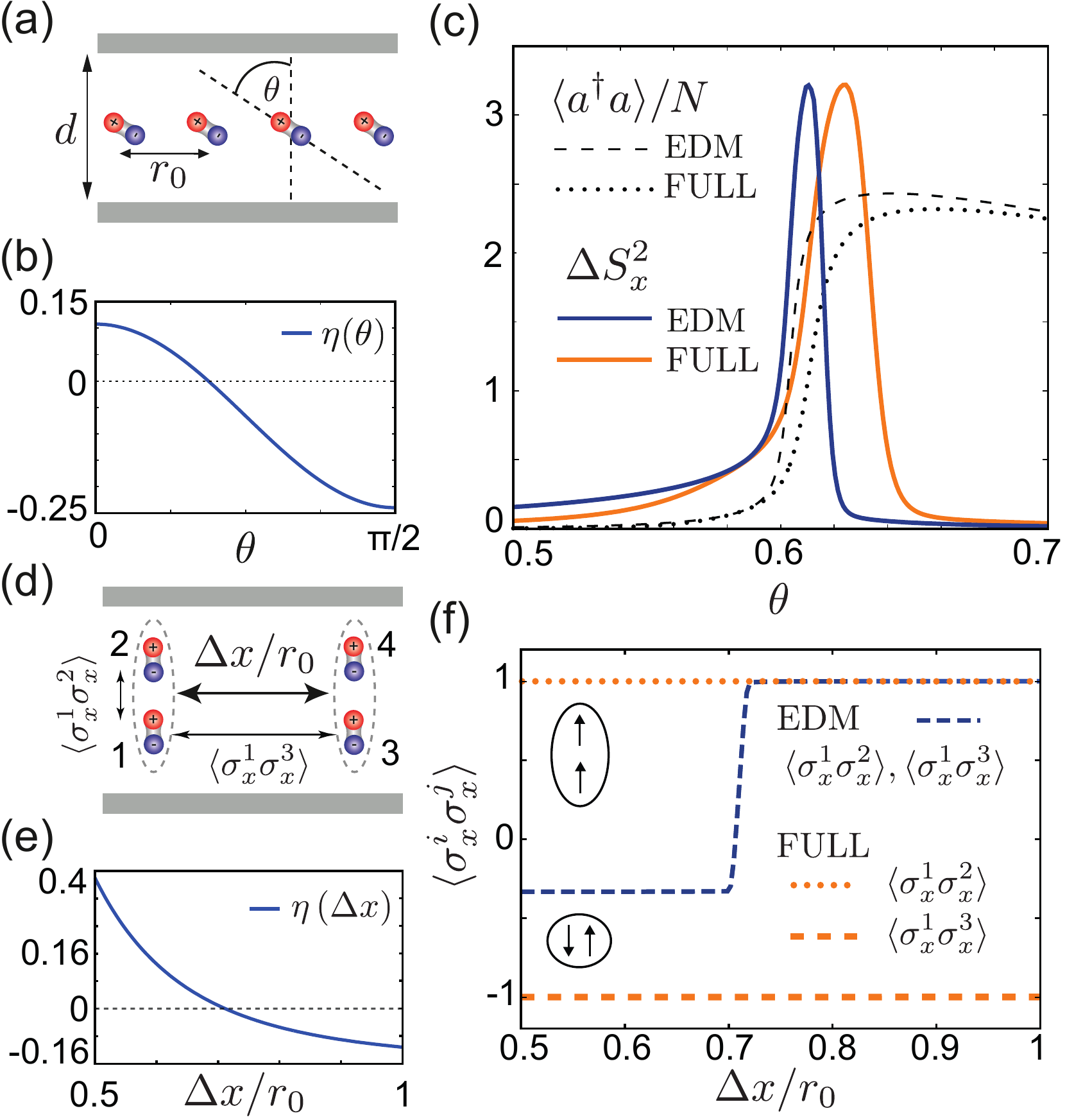}
      \caption{Comparison between the EDM and full cavity QED Hamiltonian $H_{\rm cQED}$. (a) Sketch of a setup with $N=4$ dipoles, where the sign of the dipole-dipole interactions is varied by tilting the dipoles by an angle $\theta$. The resulting averaged interaction parameter $\eta$ entering in $H_{\rm EDM}$ is plotted in (b) as a function of $\theta$.  (c) The dipole fluctuations $\Delta S_x^2$ and the photon number $\langle a^{\dag} a \rangle$ are evaluated for the ground state of the full Hamiltonian and for ground state of the corresponding EDM for different $\theta$.
In this plot $g/\omega_c=2$, $\omega_c=\omega_0$, $d=3r_0$ and $\nu=1/(4\pi)$. (d) Sketch of a setup, where two pairs of dipoles are separated by $\Delta x$. For this configuration the parameter $\eta$ entering the EDM is plotted in (e) as a function of $\Delta x$. The sign of $\eta$ changes at a value of around $\Delta x/r_0\approx 0.7$. 
(f) Plot of the ground state correlations $\langle \sigma_x^i\sigma_x^j\rangle$ evaluated for the configuration shown in (d) using the EDM and the full model.
In all numerical simulations a symmetry-breaking bias field,  $H_{\rm bias}=\lambda S_x$, where $\lambda/\omega_c=10^{-4}$, has been added to the bare Hamiltonians $H_{\rm EDM}$ and $H_{\rm cQED}$.}
      \label{Fig7FullModel}
\end{figure}

In the first scenario shown in Fig.~\ref{Fig7FullModel}(a) the dipoles are arranged in a line along the $x$-direction, but tilted by an angle $\theta$ with respect to the $z$-axis. This slightly reduces the coupling to the cavity field, but changes the dipole-dipole interactions from repulsive to attractive at a tilting angle of about $\theta\approx 0.6$.  In this case the short-range nature of the interactions is taken into account, but the sign of all the nearest-neighbor interactions is the same. For this configuration we evaluate the exact coupling matrix $\mathcal{D}_{ij}(\theta)$ (including image charges) and simulate the resulting full Hamiltonian $H_{\rm cQED}$ for different $\theta$. We also construct the corresponding extended Dicke model by replacing $g \longmapsto g\cos(\theta)$ and $\eta \longmapsto \eta(\theta)/\cos^2(\theta)$, where $\eta(\theta)$ is plotted in Fig.~\ref{Fig7FullModel}(b). In Fig.~\ref{Fig7FullModel}(c) we compare the results from the full model and the corresponding $H_{\rm EDM}$ as we tune the system from repulsive to attractive dipole-dipole interactions. We see that apart from small quantitative differences, the qualitative features of the subradiant-to-superradiant transition are in very good agreement.   

In a second scenario shown in Fig.~\ref{Fig7FullModel}(d) we consider two pairs of dipoles placed on top of each other at a fixed distance $r_0$, but with a varying separation $\Delta x$ along the $x$-direction. In this case there is a certain distance, $\Delta x\approx 0.7 r_0$, where the attractive interactions along $z$ balances the repulsive interactions along $x$ and the parameter $\eta$ changes from a positive to a negative value [cf. Fig.~\ref{Fig7FullModel}(e)]. From a naive application of the collective spin model $H_{\rm EDM}$ we would obtain around this point a transition into a superradiant phase. However, as shown in Fig.~\ref{Fig7FullModel}(f), for the ground state of the full cavity QED Hamiltonian $H_{\rm cQED}$ this is not the case and it remains subradiant. As indicated by the values of the dipole correlators $\langle \sigma_x^i\sigma_x^j\rangle$, this can be understood from the fact that the two dipoles on  top of each other align and simply form a collective spin-1 particle. The two effective spin-1 dipoles then anti-align in order to minimize the remaining attractive dipole-dipole interactions as well as the collective coupling terms. From this basic example we expect that in general  the formation of subradiant rather than superradiant ground states is more likely to occur.

\section{Conclusions}\label{sec:Conclusions}

In summary, we derived a minimal model for cavity QED, which is applicable in the regime where the coupling between a single dipole and the field mode is comparable to the bare photon energy. We discussed the physical parameters which are required to achieve this condition in a generic setup of dipoles coupled to the electric field of a lumped element resonator. This setting also permitts a natural reinterpretation of the resulting dipole-field interactions in terms an enhanced finestructure constant $\alpha$.  For $\alpha\ll 1$ our model differs from other commonly used cavity QED models mainly by the full treatment of direct dipole-dipole interactions, which, however, is most crucial for the correct prediction and interpretation of superradiant instabilities. For $\alpha\gtrsim 1$ the hybridization of individual dipoles and photons becomes relevant and leads to strong renormalization of the dipole frequency and a  cavity-induced anti-ferromagnetic ordering. This mechanism favors highly entangled subradiant ground states, where the dipoles are almost decoupled from the field.

While the analysis in this work was deliberately based on many idealizations and approximations, the general findings are applicable for a large range of different physical realizations of cavity QED systems. In traditional settings with atoms in optical cavities the effects described in this work are not directly accessible, since $\alpha\ll1$ and also the required densities for superradiant instabilities are so high that the cavity QED physics is masked by solidification and other short-range interaction effects~\cite{Griesser2016}. For organic molecules or intersubband transitions in quantum wells the value of $\alpha$ is still small, but ultrastrong collective couplings, $G\sim \omega_c,\omega_0$, and dipole-dipole interactions of similar strength become possible. In these systems the interplay between dipole-field and direct dipole-dipole interactions could be explored in more detail by using differently structured samples, which either favor or suppress ferroelectric order. This  creates an interesting connection between traditional studies of ferroelectric systems in confined geometries \cite{Bratkovsky2008}, and the dynamical USC effects explored in cavity QED.  

A value of $\alpha\sim 1$ can in principle be reached with superconducting Cooper pair boxes or electrons in gate-defined quantum dots when coupled to an $LC$ circuit with high impedance $Z\gg Z_0$~\cite{Devoret2007,Jaako2016,Bosman2017,Manucharyan2017}. Such values are possible using superinductors~\cite{Manucharyan2009}, where $Z\sim R_Q\approx 26$ k$\Omega$ can become comparable to the resistance quantum $R_Q$. Even higher values of $\alpha > 1$ can be achieved with flux-coupled circuit QED systems, where a more favorable scaling $g/\omega_c \sim \sqrt{R_Q/Z}$ is obtained~\cite{Devoret2007}. While our analysis has been restricted to electric systems, also the underlying equations of motion for such flux-coupled circuits can be cast into the form of Eqs.~\eqref{eq:EOMCircuit} and~\eqref{eq:EOMDipoles} and studied within the same theoretical framework. For example, a serial coupling of flux qubits as considered in Ref.~\cite{Jaako2016} corresponds to $\mathcal{D}_{ij}=0$ and a subradiant ground state is found.  For the parallel coupling considered in Ref.~\cite{Bamba2016} we obtain $\mathcal{D}_{ij}=\mathcal{D}<0$ and a superradiant ground state is expected. Therefore, also quite abstract circuit geometries can be reinterpreted in terms of interacting dipoles and described by $H_{\rm cQED}$.

Finally,  let us emphasize that there are already many quantum simulation platforms available, where $H_{\rm cQED}$ or $H_{\rm EDM}$ could potentially be implemented as effective models~\cite{Dimer2007}. For example, Rabi- and Dicke models are currently studied with cold atoms~\cite{Esslinger,KlinderPNAS2015,Schneeweiss2017} and trapped ions~\cite{Lv2017,SafaviNaini2017}, or using digital quantum simulation schemes with superconducting qubits~\cite{Langford2016}. Similar techniques can be used to engineer the additional $S_x^2$ terms required for the simulation of $H_{\rm EDM}$. In Ref.~\cite{Schneeweiss2017} it has been discussed that such a collective spin term even appears for a single trapped Rubidium atom, when its motion is coupled to Zeeman sublevels of the hyperfine manifold via fictitious magnetic fields. For such effective models there are in principle no constraints on the achievable parameter range and all the different regimes of cavity QED can be explored.

\acknowledgements
We thank Juraj Darmo, Karl Unterrainer and Philipp Schneeweiss for many stimulating discussions and feedback on the manuscript. This work was supported by the Austrian Science Fund (FWF) through the SFB FoQuS, Grant No. F40, the DK CoQuS, Grant No. W 1210, and the START Grant No.  Y 591-N16.

\appendix
\section{Dipole-dipole interactions in the presence of metallic plates}\label{app:DipoleDipole}
To calculate dipole-dipole interactions in the presence of the capacitor plates we follow a standard approach \cite{Takae2013} and solve the Poisson equation $\nabla^2 \phi(\vec{r}) = -\rho(\vec{r})/\epsilon_0$, where $\phi(\vec r)$ is the potential and $\rho(\vec r)$ is the charge distribution of the dipole ensemble, for metallic boundaries at $z=0$ and $z=d$ and with periodic boundary conditions in the $(x,y)$-plane (with small differences, the same calculation also holds in the case of a planar capacitor of infinite size). This allows us to account for the overall dependence of system parameters on the area $A=L^2$ and separation $d$, while avoiding a detailed numerical simulation of the field distribution near the edges of the capacitor.  

Here we consider the more general case in which all the dipoles are tilted by an angle $\theta$ with respect to the $z$ axis.
The dipole displacement is thus given by $\vec{\xi}_i = \xi_i \vec{u}_d$, where $\vec{u}_d = (\sin(\theta),0,\cos(\theta))$.
For the evaluation of the $\mathcal{D}_{ij}$ we need to calculate the field along the direction of each dipole $E(\vec r_i)=-\nabla \phi(\vec r_i)\cdot \vec{u}_d$ produced by a single dipole located at position $\vec r_j$ with charge distribution $\rho(\vec{r}) = q \xi_j \vec{u}_d \cdot \nabla \delta(\vec{r} - \vec{r}_j )$.  The general result can be written as 
\begin{equation}
E(\vec r)=    -\frac{q}{\epsilon_0}  \nabla \left[ \vec{\xi}_j \cdot \nabla_j \left( G( \vec{r}, \vec{r_j} ) \right) \right] \cdot \vec{u}_d,
\end{equation}
where $G( \vec{r}, \vec{r}' )$ is the Green's function satisfying $\nabla^2 G(\vec{r}, \vec{r}') = - \delta^{(3)}(\vec{r} - \vec{r}')$. This equation can be solved by introducing the Fourier transform $G_{\vec k}(z,z')= 1/(2\pi)^2\int G( \vec{r}, \vec{r}' ) e^{i(k_xx+k_y y)}  dxdy$, where $\vec k=(k_x,k_y)$. For the boundary conditions specified above we obtain
\begin{equation}
G_{\vec k} (z, z') = g_k(z,z')\Theta(z-z') + g_k(z',z)\Theta(z'-z).
\end{equation}
Here $k=|\vec k|$, $\Theta(z)$ is the \emph{Heaviside} step-function and 
\begin{equation}\label{eq:dipoleGreenFourier}
\begin{split}
 &g_k(z,z') = \frac{e^{-k(z-z')}}{2k}
\\
& - \frac{\sinh(kz) e^{-k(d -z')} + \sinh(k(d - z))e^{-kz'} }{2k\sinh(k d)}.
\end{split}
\end{equation}
From this result we can immediately evaluate the total induced charge on the capacitor plates
$
Q_{\rm in} = \int_A dxdy ~ \sigma_{\rm in}(x,y) = \int_A dxdy  ~ \epsilon_0 \vec{e}_z\cdot\vec{E}(x,y,z=d).
$
 It follows that
$
Q_{\rm in}= - q\sum_{i=1}^N  \xi_i \cos(\theta) \partial_z    \partial_{z_i}  G_{k=0}(z=d, z_i).
$
To evaluate $G_{k=0}$ we use
\begin{equation}
\lim_{k \rightarrow 0 } G_k(z, z_i) = \frac{(z+z_i -|z-z_i| )}{2} - \frac{z z_i}{d},
\end{equation}
and we obtain $Q_{\rm in}= -q \cos(\theta) \sum_i \xi_i/d$.

The full electric field in real space can be reconstructed by an inverse Fourier transform of Eq. \eqref{eq:dipoleGreenFourier} (which can be performed in the cases of finite size and periodic boundary conditions or infinite plane with the field vanishing fast enough at infinity).
The explicit expression of the local field in the case of a finite system with periodic boundary conditions is
\begin{equation}
\begin{split}
\vec{E}(\vec r) =- \frac{q }{4\pi \epsilon_0}\sum_{ \vec{m}\in \mathbb{Z}^3} \nabla \cdot \nabla_j \left( \frac{\vec{\xi}_j}{\left| \vec{r} - \vec{r}_j -\vec{h} \right|} + \frac{\vec{\xi}_{*j}}{\left| \vec{r} - \vec{r}_{*j} -\vec{h} \right|} \right)
\end{split}
\label{local_field_general_PBC}
\end{equation}
where $\vec{h}=(Lm_x, L m_y, 2d m_z)$, $\vec{\xi}_{*j} =(-\xi_j^x, -\xi_j^y, \xi_j^z)$ and $\vec{r}_{*j} = (x_j, y_j, -z_j)$. This compact expression is nothing else than the field generated by the $j$-th dipole, plus the field generated by the infinite images of each dipole reflected by the metallic boundaries along $z$, plus the field generated by the infinite copies of the system because of the periodic boundaries along $(x,y)$.
The same result holds for the infinite capacitor, with the only difference that the summation over infinite copies of the system disappears.
Using the above definition of the tilted dipole moment, and considering the case of an infinite planar size capacitor, we obtain
\begin{equation}
\mathcal{D}_{ij} =  \mathcal{D}_{ij}^0+\sum_{n \neq 0} (F^{n,S}_{ij} + F^{n,O}_{ij}  ).
\end{equation}
Here $\mathcal{D}_{ij}^0$ is the result given in Eq.~\eqref{eq:Dij} in the absence of boundaries and 
\begin{equation}
\begin{split}
F^{n, S}_{i,j}   &= r_0^3 \left( \frac{1}{|\vec{r}_i -\vec{r}_{*j} - \vec{h}_n|^3} \right.
\\
 &  \left. -  \frac{3( (z_i - z_j + 2dn)\cos(\theta) + (x_i - x_j)\sin(\theta))^2}{|\vec{r}_i -\vec{r}_{*j} - \vec{h}_n|^5} \right) ,
 \\
F^{n, O}_{i,j}  &= r_0^3 \left( \frac{1}{|\vec{r}_i -\vec{r}_{*j} - \vec{h}_n|^3} \right.
\\
 &  \left. - \frac{3( (z_i + z_j + 2dn)^2\cos(\theta)^2 - (x_i - x_j)^2\sin(\theta)^2)}{|\vec{r}_i -\vec{r}_{*j} - \vec{h}_n|^5} \right) ,
\end{split}
\end{equation}
where $\vec{h}_n = (0,0,2dn)$.

\section{Polariton modes}\label{app:Polaritons}
For harmonically bound dipoles the general set of equations \eqref{eq:EOMCircuit} and \eqref{eq:EOMDipoles}  can be solved by decomposing the dipole variables as $
\xi_i=  \sum_{n} c_n(i)  \mathcal{Z}_n$, 
where the $c_n(i)$ are normalized eigenmodes of the dipole-dipole interaction matrix, which obey
\begin{equation}
\eta_n c_n(i)  - \sum_{j} \mathcal{D}_{ij} c_n(j) =0.
\end{equation}
By introducing dimensionless variables $\tilde{\Phi} = \Phi / \bar{\Phi}$ and $\tilde{\mathcal{Z}}_n = \mathcal{Z} / \bar{\mathcal{Z}}$, where  $\bar{\Phi}^2 =  \sqrt{L/C}m \omega_0 \bar{\mathcal{Z}}^2 $ and  $\bar{\mathcal{Z}}$ is an arbitrary length scale,
we obtain the set of coupled equations 
\begin{equation}
\begin{split}
\ddot{\tilde{\mathcal{Z}}}_n +   \left(\omega_0^2 + \eta_n \omega_{\rm p}^2\right)  \tilde{\mathcal{Z}}_n   = -\omega_{\rm p}\sqrt{\frac{\omega_0}{\omega_c} \nu_n}\dot{\tilde{\Phi}} , 
\\
\qquad   \ddot{\tilde{\Phi}} +\omega_c^2 \tilde{\Phi} =   \sum_n \omega_{\rm p}\sqrt{\frac{\omega_c}{\omega_0} \nu_n} \dot{\tilde{\mathcal{Z}}}_n.
\end{split}
\end{equation}
Here we defined the parameters $ \nu_n = r_0^3 \left[\sum_i c_n(i)\right]^2 /V $, which characterize the relative coupling strength between the resonator and each dipole mode. In the limit where the resonator is dominantly coupled to a single collective mode, i.e., $\nu_0\simeq \nu $ and $\nu_{n\neq 0}\simeq 0$, we recover Eqs. \eqref{eq:EQM_Z} and  \eqref{eq:EQM_Phi}. The resulting eigenvalue equation is given by 
\begin{equation}\label{eq:EigenvaluesApp}
(\Omega^2-\omega_c^2)\prod_n (\Omega^2-\omega_n^2) \left[ 1- \sum_n  \frac{\nu_n \omega_{\rm p}^2 \Omega^2}{(\Omega^2-\omega_c^2)(\Omega^2-\omega_n^2)}\right]=0, 
\end{equation}
where $\omega_n^2 = \omega_0^2 + \eta_n \omega_{\rm p}^2$. 
Since we are interested in the spectrum of coupled modes, we can assume $\Omega \neq \omega_c, \omega_n$ and look for the solutions of
\begin{equation}
\sum_n \frac{\omega_{\rm p}^2 \nu_n}{\Omega^2 - \omega_n^2} = 1 - \frac{\omega_c^2}{\Omega^2}.
\label{coupled_harmonic_eigen_equation}
\end{equation}
The appearance of an unstable mode is indicated by a solution of this equation for which  $\Omega\rightarrow 0$. This is only possible if one of the mode frequencies of the dipole ensemble vanishes, i.e., for $\omega_n\rightarrow0$.
 
\section{Single-mode approximation in the electric dipole gauge}\label{app:SingleModeHopfield}
Starting from Hamiltonian~\eqref{eq:HamHopfield} in the electric dipole gauge, we can decompose the operator for the displacement field into a set of orthogonal modes,
\begin{equation}
\vec{D}(\vec{r}) =\sum_{k} \sqrt{\frac{\hbar \omega_k\epsilon_0}{2}}\left[a_{k}  \vec{f}_{k}(\vec{r}) +   \vec{f}^*_{k}(\vec{r}) a_{k}^\dag\right],
\end{equation}
where the $a_k$ ($a_k^\dag$) are annihilation (creation) operators for a mode of frequency $\omega_k$ and mode function $\vec{f}_{k}(\vec{r})$. The $\vec{f}_{k}(\vec{r})$ are solutions of the Helmholtz equation for the geometry under consideration and they are normalized to $\int d^3 r  \vec{f}_{k}^*(\vec{r}) \vec{f}_{k'}(\vec{r})= \delta_{k,k'}$. Using this decomposition the Hamiltonian reads
\begin{equation}
\begin{split}
H_{\rm HM} &= \sum_{k} \hbar \omega_k a^{\dag}_{k} a_{k} + H_{\rm matter} + \frac{1}{2\epsilon_0}\int d^3r P^2(\vec{r})
\\
&+ \sum_{k}  \int d^3r  \sqrt{\frac{\hbar \omega_{k}}{2\epsilon_0}} \left[ a_{k} \vec{f}_{k}(\vec{r})\cdot \vec{P}(\vec{r}) +{\rm H.c.}\right],
\end{split}
\end{equation}
which at this stage is still exact. 

We are now interested in the situation where only the lowest frequency mode is resonant with the dipoles, i.e. $\omega_{k_0}\equiv\omega_c \approx \omega_0$, while all other modes are far separated in frequency, $\omega_{k}\gg\omega_0$. By looking, for example, at the Heisenberg equations of motion for these modes,
\begin{equation}
\partial_t a_{k} = - i \omega_k a_{k} - i\sqrt{\frac{\omega_k}{2\hbar\epsilon_0}}\int d^3 r \vec{f}_{k}(\vec{r}) \cdot \vec{P}(\vec{r}),
\end{equation}
we can adiabatically eliminate their dynamics by approximating $a_{k \neq k_0}(t) \simeq - \sqrt{\frac{1}{2\hbar \omega_k\epsilon_0 }}\int d^3r \vec{f}_{k}(\vec{r}) \cdot \vec{P}(\vec{r},t) $. The resulting dynamics for the dipoles and the remaining cavity mode can then be modelled by an effective low-frequency Hamiltonian 
\begin{equation}\label{eq:ham_SMH}
\begin{split}
H_{\rm SMH} &= \hbar \omega_c a^{\dag}_c a _c + H_{\rm matter} +H_{\rm dd}  
\\
&+ \sqrt{\frac{\hbar \omega_c}{2\epsilon_0}}(a_c + a_c^{\dag})\int d^3r \vec{f}_{c}(\vec{r}) \cdot \vec{P}(\vec{r}) \\&+ \frac{1}{2\epsilon_0}\left[ \int d^3r \vec{f}_{c}(\vec{r}) \cdot \vec{P}(\vec{r}) \right]^2,
\end{split}
\end{equation}
where we replaced the index $k_0$ by the index $c$ to be consistent with the notation used in the main text.
Here we have introduced the dipole-dipole interaction term 
\begin{equation}
H_{\rm dd} = \frac{1}{2\epsilon_0}\sum_{i,j}\int d^3r d^3r' P_i(\vec{r})K_{ij}(\vec{r},\vec{r}')P_j(\vec{r}'),
\end{equation}
where $K_{ij}(\vec{r},\vec{r}') = \delta_{ij}\delta(\vec{r} - \vec{r}') - \sum_{k} f_{k,i}(\vec{r})f_{k,j}(\vec{r}')$. Note that the sum in $K_{ij}$ runs over \emph{all} modes. To compensate for the $k_0$ term, which has not been adiabatically eliminated, the nonlocal $P^2$-term in Eq. \eqref{eq:ham_SMH} has been introduced.  

Equation \eqref{eq:ham_SMH} represents a generic \emph{single-mode} version of the Hopfield model in the dipole gauge (for a similar calculation in the Coulomb gauge see Ref. \cite{Tufarelli2015}). It shows that high frequency modes cannot be just omitted, but they contribute in second-order perturbation theory to relevant interactions terms between the dipoles. To illustrate this point let us consider the limiting case $V\rightarrow \infty$, where the mode functions are plane waves,
\begin{equation}
\vec f_{ \vec k, \lambda}(\vec{r}) = \vec \varepsilon_\lambda(k)\frac{1}{\sqrt{V}} e^{i\vec{k}\cdot \vec{r}},
\end{equation}
labeled by the wavevector $\vec k$ and the polarization vector $\vec \varepsilon_{\lambda=1,2}\perp \vec k$.
The kernel matrix for the complex mode functions gives us the dipole-dipole interaction
\begin{equation}
\begin{split}
&K_{ij}(\vec{r},\vec{r}')  =  \delta_{ij}\delta(\vec{r} - \vec{r}') - \frac{1}{V}\sum_{\vec k ,\lambda} \varepsilon^\lambda_i(\vec k) \varepsilon^\lambda_j(\vec k)e^{i\vec{k}\cdot (\vec{r} - \vec{r}')}
\\
 &= \frac{1}{4\pi} \left[ \frac{\delta_{i,j}}{\left| \vec{r} - \vec{r}' \right|^3} - \frac{3 (\vec{r} - \vec{r}')_i(\vec{r}-\vec{r}')_j}{\left|\vec{r} - \vec{r}' \right|^5} \right] - \frac{1}{3}\delta(\vec{r} - \vec{r}'),
\end{split}
\end{equation}
where we made use of the transversality of the electro-magnetic field~\cite{PhotonsAndAtoms} and replaced the sums over the $k$-vectors by integrals. For a finite volume and metallic boundaries, a similar calculation would reproduce the modified dipole-dipole couplings $\sim \mathcal{D}_{ij}$, as used in our model.

\end{document}